\documentclass[a4paper,12pt]{article}

\usepackage{amsmath}
\usepackage{mathtools}
\usepackage{physics}
\usepackage[utf8]{inputenc}
\usepackage{amssymb}
\usepackage{authblk}
\usepackage{graphicx,amssymb,amsfonts}
\usepackage[sort&compress,numbers]{natbib}
\usepackage{float}
\usepackage{xcolor}
\newcommand\be{\begin{equation}}
\newcommand\ee{\end{equation}}
\newcommand\bea{\begin{eqnarray}}
\newcommand\eea{\end{eqnarray}}
\newcommand\del{\partial}

\title{Dyonic Taub-NUT-AdS: \\
Unconstraint Thermodynamics and Phase Structure}
\author[a,b]{Adel Awad \thanks{a.awad@sci.asu.edu.eg}}
\author[a]{Esraa Elkhateeb\thanks{dr.esraali@sci.asu.edu.eg}}

\affil[a]{\footnotesize \it Department of Physics,
Faculty of Science, Ain Shams University, Cairo 11566, Egypt}
\affil[b]{\footnotesize \it Centre for Theoretical Physics, the British University in Egypt, El Sherouk City 11837, Egypt}
\date{}

\begin{document}
\maketitle
\begin{abstract}
Here we extend the approach developed in \cite{adel_2} to study the thermodynamics of Taub-NUT-AdS and dyonic Taub-NUT-AdS solutions. Furthermore, we investigate in details the possible phase structures of the dyonic Taub-NUT-AdS solution. We show that the first law, Gibbs-Duhem and Smarr's relations are all satisfied for both solutions. Our study of phase structures shows some intriguing features, which were not reported before, among which the existence of two distinguished critical points with a region of continuous phase transitions in between, and the possibility of merging them into one point. To analyze these phases we consider both canonical and mixed ensembles. The two distinguished critical points occur for the canonical case as well as the mixed cases with  $1/2 \le \phi_e < 1$. Another interesting case is the mixed ensemble with $\phi_e \ge 1$, where we have one critical point but the continuous phase transition region in the $P-T$ diagram is close to the origin, in contrast with what happens in Reissner-Nordstrom-AdS solutions and Van der Waals fluids, i.e., the continuous phase transition happens only for low enough pressures and temperatures!
\end{abstract}
\section{Introduction}
Taub-NUT solution was first introduced by Taub in \cite{taub}, then was investigated by Newman, Unti, and Tamburino (NUT) in \cite{NUT}. It is a vacuum solution with two Killing vectors, which possess a conical singularity. This singularity forms a string called Misner string \cite{misner}, which is considered to be the gravitational analog of Dirac' string. This spacetime and its thermodynamics have been widely studied in the Euclidean section in literature, see \cite{hunter,hunter+hawking,page+hunter+hawking,clifford98} and references therein. To render this conical singularity invisible, a restriction must be imposed on the time periodicity $\beta$ which leads to a relation between the nut parameter $n$ and the horizon radius $r_0$, namely, $\beta(r_0)=8\pi n$. As a result, there is no independent work term, or $\phi_n dn$, for the nut charge in the first law similar to that of Kerr, $\Omega dJ$, or charged solution, $\phi dQ$. In such thermodynamics, the entropy is not equal to a quarter of the horizon area, the temperature is not well defined in the limit $n \rightarrow 0$, and the thermodynamic volume can be negative! 

In recent works, several authors \cite{Durka,mann_1,Geom_inter,adel_1,adel_2,ww,smarr,Nutty_dyons_1,nutty_rot,rot_nuts,rot_tn} considered the possibility of constructing unconstrained thermodynamics for Taub-NUT spaces in the Lorentzian section by relaxing the above time-periodicity condition and obtaining an independent work term which depends on $n$. These authors found that in order to formulate a full cohomogeneity first law where the nut parameter can vary independently, it is natural to introduce a new charge ${\cal N}$ (which vanishes as we send $n$ to zero) together with its conjugate chemical potential $\psi$. Then the first law can be written as $dU = TdS+VdP+\psi d{\cal N}$, where the entropy is equal to a quarter of the horizon area and the temperature is well defined in the limit $n \rightarrow 0$. There are several proposals in the literature which carry this spirit but they have different conjugate pairs $(\psi,{\cal N})$. Another quantity that might not be the same in these proposals is the internal energy $U$. 

The leading approach for unconstrained thermodynamics was introduced in \cite{mann_1} where the authors proposed a conjugate pair $(\psi,{\cal N})$ to show the realization of the first law and the entropy as the area of the horizon. A geometric interpretation of this approach was presented afterwards in \cite{Geom_inter}, where $\psi$ is shown to be proportional to Misner string temperature, and ${\cal N}$ can be interpreted as its entropy (Obtained from a Komar-type integration over Misner tubes). Now we have a multi-temperature system with a horizon and a Misner temperatures. This reduces the cohomogeneity of the first law since at equilibrium both temperatures should be the same which takes us back to the restriction $\beta(r_0)=8\pi n$. In this proposal, it is unclear if the nut charge ${\cal N}$ is conserved or not, which might affect the validity of the first law. This work inspired several authors \cite{adel_1,adel_2,ww,smarr,rot_tn} to further study this approach creating different possible ways to study this class of solutions with nut charge.

In this work, we propose another unconstrained thermodynamics approach for Taub-NUT-AdS spaces, following our earlier works \cite{adel_1,adel_2} on Taub-NUT solutions in Minkowski space. Here we introduce a nut charge $N=n\,(1+4n^2/L^2)$ and its potential $\phi_N$, where the charge $N$ is conserved since it is the dual mass obtained from Komar's integral. We will see that one can also work with the conjugate pair $(n,\phi_n)$ instead of $(N,\phi_N)$ since $\phi_n$ is related to $\phi_N$ by some factor. When this approach was applied to Taub-NUT in flat space \cite{adel_2}, the resulting internal energy is not the mass of spacetime but $U=M-n\phi_n$. In this work we show that in extended thermodynamics, i.e., allowing AdS radius to change, the enthalpy for Taub-NUT-AdS, is $H=M-n\phi_n$. This mismatch between the enthalpy and mass is a key ingredient in our construction; The enthalpy is not identified with the gravitational mass anymore, instead, it is related to the mass by a Legendre transform which vanishes as we send $n$ to zero. This is similar to the $PV$ term that appears in the known $U-H$ Legendre transformation commonly used in extended thermodynamics, namely, $H = U + PV$. Here we extend our idea to the charged dyonic Taub-NUT solutions in AdS to build a consistent thermodynamics with both electric and magnetic charges appearing in the first law. We show that the first law, Gibbs-Duhem, and Smarr's relations are all satisfied. Also, the entropy is the area of the horizon and the temperature goes to that of dyonic-AdS black holes as $n\rightarrow 0$. Furthermore, we have studied phase structures in detail using the above approach which shows some new interesting features which were not reported elsewhere. In particular, we found two distinguished critical points, between them there exists a continuous phase transition region. We also studied in details the possibility of merging these two points into one in the canonical and mixed ensembles. As we will see below we have studied the phase structure of two ensembles; the canonical ensemble in which the electric potential is set $\phi_e=0$, and the mixed ensemble in which the magnetic charge is set $q_m=0$. The canonical case and a particular class of the mixed case have the two distinguished critical points. Another intriguing case is the one with $\phi_e \ge 1$ in the mixed ensemble which has one critical point but the continuous phase transition region in the $P-T$ diagram is close to the origin in contrast to the usual case of the charged black holes in AdS. Also, the continuous phase transition occurs in this case, if we go to low enough pressure and temperature.

Our paper is organized as follows; in section (2) we study neutral Taub-NUT-AdS in extended thermodynamics where we calculate various thermodynamic quantities and show that the first law, Gibbs-Duhem, and Smarr's relations are all satisfied. In section (3) we study Dyonic-Taub-NUT-AdS extended thermodynamics through calculating its thermodynamic quantities and show again that the first law, Gibbs-Duhem, and Smarr's relations are all satisfied. In section (4) we divide our study into canonical and mixed ensembles where we show the existence of two critical points. We further divide the mixed cases into a few sub-cases for further investigation. In section (5) we present our conclusion with some remarks on possible extensions of this work to other solutions with nut charges. 
%%%%%%%%%%%%%%%%%%%%%%%%%%%%%%%%%%%%%%%%%%%%%%%%%%%%%%%%%%%%%%%

\section{Taub-NUT-AdS Space Thermodynamics}
It is constructive to discuss first the neutral Taub-Nut-AdS case to show some features of its thermodynamics. The metric of this spacetime is given by \cite{page+hunter+hawking, Clement1}
\begin{equation}
    ds^2=-f(r)\,(\,dt-2n(\cos{\theta}+k)\, d\phi\,)^2\,+\,\frac{dr^2}{f(r)}\,+\,(r^2+n^2)\,(\,d\theta^2+\sin^2{\theta}\,d\phi_e^2), \label{metric}
\end{equation}
where, $r$ is the radial coordinate, $\phi$, and $\theta$ are the spherical polar coordinates angles, and $n$ is the nut parameter. The function $f(r)$ is given by
\begin{equation}
    f(r)\,=\,\frac{r^2-n^2-2\,m\,r}{r^2+n^2} \, + \, \frac{r^4+6\,n^2\,r^2-3\,n^4}{\left(r^2+n^2\right)\,L^2}. \label{solnf1}
\end{equation}
Here $m$ is the mass parameter and $L$ is the AdS radius which is related to the cosmological constant through $\Lambda = -\frac{3}{L^2}$.

The parameter $k$ is a dimensionless parameter which determines the position of Misner string, was introduced in \cite{Manko} (see also \cite{Durka,Clement1} for some discussion on it). In particular, for $k=+1$, a single Misner string exists along the positive z-axis, while, for $k=-1$, a single string exists along the negative z-axis. But if $k=0$, the two strings exist symmetrically along the z-axis, with a conical singularity along the z-axis as well. Imposing the periodicity condition $\beta= 8\pi n$ leads to the removal of this conical singularity, with the cost of producing closed timelike curves in this spacetime \cite{Clement1,Clement2}. As a result, thermodynamic properties of such spacetime possess peculiar properties. For example, the entropy is not the area of the horizon, the temperature is fixed by the parameter $n$, and the temperature does not reduce to that of Schwarzschild-AdS as $n\rightarrow 0$, in fact, it is not well defined in this limit.

\subsection{Thermodynamics}
The thermodynamics of the above solution is characterized by a horizon temperature which takes the form 
\begin{equation} \label{tempTN}
  T=\frac{L^2 + 3\, \left(n^2+r_0^2\right)}{4 \pi\,  L^2\, r_0}
\end{equation}
Using counter-term method \cite{bala,surf99} one can calculate the finite on-shell gravitational action of the above solution which has the form 
\begin{equation}
   I=\beta \left( \frac{m}{2} - \frac{r_0^3 + 3 n^2 r_0}{2\,  L^2} \right)
   \label{I_TN}
\end{equation}

Here we are following the thermodynamic treatment introduced in \cite{adel_1,adel_2} where we do not impose the periodicity condition. Here we introduce a conserved nut charge $N=n\,(1+4n^2/L^2)$, where the charge $N$ is conserved since it is the dual mass obtained from Komar's integral and $\phi_N$ is its potential as we will see below. We can also work with the conjugate pair $(n,\phi_n)$ instead of $(N,\phi_N)$, as we will see below. As a result, the entropy is the area of the horizon, while the limit $n\rightarrow 0$ reduces the temperature to that of Schwarzschild-AdS $T=f'(r_0)/4\pi$, in contrast with constrained treatments \cite{clifford98,mann06}.

 Euclidean path integral boundary conditions fix the boundary metric, which also fixes the nut charge, therefore, we have a canonical ensemble with the following partition function \be Z_{can}(\beta,n)=e^{-\beta\, F},\ee where $F$ is the Helmholtz free energy, $F=I/\beta$. The entropy is related to the area of the horizon
\be S = \beta\del_{\beta} I-I=\pi \left(r_0^2 + n^2 \right). \ee The chemical potential can be calculated from the free energy
\be \phi_n=\left(\frac{\del F}{\del n}\right)_{T}, \ee
or, 
\be \phi_N=\left(\frac{\del F}{\del N}\right)_{T}. \ee
Since AdS radius $L$ is a fixed parameter here, working with $N$ or $n$ are the same since $\phi_N$ and $\phi_n$ are related,
 \be \phi_N={L^2 \over L^2+12n^2} \phi_n.\ee
\begin{equation} \label{chem}
    \phi_n=\frac{3\,n\,\left(r_0^2-n^2\right) -n\, L^2}{2\,r_0\,L^2}.
\end{equation}

To calculate the mass one can use the generalized Komar's integral introduced for asymptotically anti-de Sitter solutions in \cite{kastor, Geom_inter}. The mass of the Taub-Nut solution is given by
\be M= -{1 \over 4 \pi}\int_{S_{\infty}^2}\, (^{*}d\xi+{2} \Lambda \omega)= m, \label{mass} \ee
where $\xi=\del_t$ is a time-like Killing vector and $\omega$ satisfies $\xi^{\mu}=\nabla_{\nu}\omega^{\nu\mu}$, where 
\be \omega =  {\frac{2 \; n}{3}}\,dr\wedge(dt-2n\cos{\theta}\, d\phi)- { \frac{2r}{3}}\,(r^2+n^2)\sin{\theta}\, d\theta\wedge d\phi.\ee
Also,
\be *d\xi=- {2 \; n \over r^2+n^2 }f\,dt\wedge dr-{4n^2\, f  \over r^2+n^2 }\cos{\theta}\,dr\wedge d\phi -f'(r^2+n^2)\sin{\theta}d\theta \wedge d\phi. \ee

As was pointed out in \cite{adel_2}, these solutions are not trivial in the sense that there are mass distributions along the Misner string. To see that one can calculate these mass distributions using the above Komar's integral for $S^2_{h}$ as well. 
Calculating the mass contained in $S^2_{\infty}$ gives $M=m$, while the mass inside the horizon is \be M^h={r_0^2+n^2\over 2 r_0}+{r_0^4+3n^4 \over 2r_0L^2}. \ee
Notice that this expression can be written as $M^h=M-2n\phi_n$. This reveals the existence of mass along the Misner string, similar to what is found in \cite{adel_2}, which is given by $M^{s}=2n\phi_n$. To see that let us calculate it along the positive z-axis, see \cite{adel_2} for details, one gets \be M^{+}= -{1 \over 4 \pi} \int_{T_+}  {^*d\xi}= n\,\phi_n.\ee For the negative-z-axis the mass calculation gives \be M^{-}= n\,\phi_n,\ee
which explains why the mass at the horizon is different from that at infinity. One can also calculate the conserved charge $N$ as the dual mass, or \be N= {1 \over 4 \pi}\int_{S_{\infty}^2}\, (d\xi-{2} \Lambda ^{*}\omega)=n\left(1+{4n^2\over L^2}\right), \label{nut} \ee
where its chemical potential is given by
\begin{equation} \label{chem1}
    \phi_N=\left({\del F\over \del N}\right)_{T} =\frac{3\,n\,\left(r_0^2-n^2\right) -n\, L^2}{2\,r_0\,(L^2+12n^2)}.\end{equation} As a result, one can see that $\phi_n$ is the same as $\phi_N$ apart from a $n$-dependent factor. Because of the previous property and to keep the analysis simple, one can use the pair $(n,\phi_n)$ instead of $(N,\phi_N)$.

The internal energy is given by
\begin{align}
    U & = -\del_{\beta}\, {\text{ln}\, Z_{can}} = \del_{\beta} I=  M\,-\,n\, \phi_n \\
    & = \frac{r_0^3 + r_0 \left(3 n^2 + L^2\right)}{2\,  L^2}.
\end{align}
The free energy is related to the action through the relation 
\begin{equation}
    F=\frac{I}{\beta} =  M\, -\,T\,S\, -\,n\, \phi_n.
\end{equation}
Accordingly
\begin{equation}
    dF= -\,S\,dT\, +\, \phi_n\,dn,
\end{equation}
and
\begin{equation}
  \left(\frac{\partial F}{\partial T}\right)_{n,P}=-S, \,\,\,\,\, \left(\frac{\partial F}{\partial n}\right)_{T,P}=\phi_n. \, \,\,\,\,
\end{equation}
The above quantities satisfy the first law of thermodynamics 
\begin{equation}
  dU=d\left(M\, -\,n\, \phi_n\right)=\,T\,dS\, +\, \phi_n\,dn. 
\end{equation}

\subsubsection{Extended Thermodynamics}
By considering a varying cosmological constant one can add a pressure to thermodynamic relations, or $P=\frac{3}{8\pi L^2}$, with a conjugate volume $V$. In this case, the gravitational action is related to Gibbs energy, $G$, rather than Helmholtz energy $F$. The variation of $G=\frac{I}{\beta}$ is given by 
\begin{equation}
    dG= -\,S\,dT\, +\, \phi_n\,dn \,+\, V\, dP,
\end{equation}
or,
\begin{equation}
    dG= -\,S\,dT\, +\, \phi_N\,dN \,+\, V'\, dP,
\end{equation}
where,
\begin{equation}
  \left(\frac{\partial G}{\partial T}\right)_{n,P}=-S, \,\,\,\,\, \left(\frac{\partial G}{\partial n}\right)_{T,P}=\phi_n, \, \,\,\,\,\,   \left(\frac{\partial G}{\partial P}\right)_{n,T}=V,
\end{equation}
and,
\be V= {4\pi r_0^3 \over 3}\,\left(1+{3n^3\over r_0^2}\right). \ee
Notice that $V'$ is different from $V$ and one can choose to work with the pair $(n,\phi_n)$ or $(N,\phi_N)$. Here to keep the analysis simple we choose to work with pair $(n,\phi_n)$ and we will do that in all the coming discussions.

The internal energy of the system and other quantities satisfy the following Smarr's relation 
\begin{equation}
    U=M-n\phi_n-PV=2\,T\,S\, +\,n\, \phi_n \,-P\, V.
\end{equation}
The first law of thermodynamics as well as the Gibbs-Duhem relation are satisfied
\begin{equation}
  dU=d\left(M\, -\,n\, \phi_n-PV \right)=\,T\,dS\, +\, \phi_n\,dn \,-\, P\, dV. 
\end{equation}
\begin{equation}
 G=U+PV-TS=M-n\phi_n-TS. 
\end{equation}
%%%%%%%%%%%%%%%%%%%%%%%%%%%%%%%%%%%%%%%%%%%%%%%%%%%%%%%%%%%%%%
\section{Dyonic Taub-NUT AdS Thermodynamics}
Now we are ready to discuss the Taub-NUT-AdS case with electric and magnetic charges and apply the thermodynamical treatment introduced in \cite{adel_2}. First, we are going to calculate the electric and magnetic charges of the solution as well as their potentials, then calculate various thermodynamic quantities and check the validity of the first law, Gibbs-Duhem, and Smarr`s relations.

\subsection{Charges and Potentials}
For this solution, the metric has the same form as the uncharged Taub-NUT case in eqn.(\ref{metric}), but the function $f(r)$ is given by
\begin{equation}  \label{fr}
    f(r)\,=\,\frac{r^2+q_e^2+q_m^2-n^2-2\,m\,r}{r^2+n^2} \, + \, \frac{r^4+6\,n^2\,r^2-3\,n^4}{\left(r^2+n^2\right)\,L^2},
\end{equation}
where $q_e$ and $q_m$ are the electric and magnetic charges. The gauge potential $A_\mu$ is given by
\begin{equation} \label{pot}
    A=\left(\frac{n\,q_m - q_e r}{r^2 + n^2} + \phi_e\right)dt + \left(\left[\frac{2\,n\,q_e r + q_m\left(r^2-n^2\right)}{\left(r^2+n^2\right)}\right]cos\theta+C\right)d\phi,
\end{equation}
where $\phi_e$ and $C$ are integration constants. The gauge potential in eqn.(\ref{pot}) and the above metric satisfy the field equations  
\begin{equation}
     G_{\mu \nu}=\kappa \,T_{\mu \nu} , \hspace{0.7 in} \nabla_\mu F^{\mu \nu} =\,0,
\end{equation}
where
\begin{equation}
   T_{\mu \nu}=F_{\mu \alpha} F_\nu^\alpha - \frac{1}{4} g_{\mu \nu} F^2.
\end{equation}
The magnetic charge in a spatial region $\Sigma$, with a boundary $\del \Sigma$, is given by \be Q'_m=-{1 \over 4\,\pi }\int_{\Sigma}\, dF=-{1 \over 4\,\pi }\int_{\del \Sigma}\, F. \ee
The magnetic flux at any radius $r$ is \be q_m(r)= -{1 \over 4\,\pi }\int_{S^2_r}\,F= \frac{q_m\,(r^2-n^2)+2\,n\,q_e\,r}{r^2+n^2},\ee
which produces a magnetic charge at radial infinity, \be Q_m^{\infty}=q_m,\ee
and a magnetic charge at the horizon \be Q_m^h=(q_m+2\,n\phi_e).\ee
The electric charge in a spacial region $\Sigma$ is given by \be Q'_e={1 \over 4\,\pi }\int_{\Sigma}\, d^*F={1 \over 4\,\pi }\int_{\del \Sigma}\, ^*F,\ee
where, $^*F$ is the Hodge dual of $F$. Also, the electric flux at any radius $r$ is \be q_e(r)= {1 \over 4\,\pi }\int_{S^2_r}\,^*F= \frac{q_e\,(r^2-n^2)-2\,n\,q_m\,r}{r^2+n^2},\ee
which produces the following electric charge at infinity \be Q_e^{\infty}=q_e,\ee
but, at the horizon it takes the form \be Q_e^{h}=(q_e-2\,n\phi_m).\ee
The electric and magnetic potentials are defined as
\be \phi_e=\Phi_e|_{\infty}-\Phi_e|_{h},\ee
\be \phi_m=\Phi_m|_{\infty}-\Phi_m|_{h}={q_m+n\,\phi_e\over r_0},\ee where $\Phi_e=A_{\mu}\xi^{\mu}$ and $\Phi_m=B_{\mu}\xi^{\mu}$, with $\xi$ is a time-like Killing vector. Also, the one-form $B$ is the solution of $dB=^*F$, which is  given by
\be B= \left(-\frac{n\,q_e\,+\,q_m\,r\,}{r^2\,+\,n^2}+V'\right)\,dt\,+\left(\frac{2\,n\,q_m\,r+q_e\,(r^2-n^2)}{r^2+n^2} + C'\right) \, \cos\theta\,d\phi , \ee
where $V'$ and $C'$ are integration constants. We will see now the importance of these integration constants.

The thermodynamics imposes certain regularity conditions on the gauge potential $A_{\mu}$ \cite{adel_2}. To have a nonsingular one-form $A$ on the horizon, the charges $q_e$, $q_m$ and the potential $\phi_e$ should be related as follows,
\begin{equation} \label{qe}
    q_e\,=\,\frac{n\,q_m\,+\,\phi_e\,(n^2\,+\,r_0^2)}{r_0}.
\end{equation}
Also, to have a nonsingular potential along the z-axis we should have two patches for $A$, one is smooth on the northern hemisphere, and the other is smooth on the southern hemisphere as in Dirac's monopole case, or
\be C_{\pm}=\mp(q_m+2n\phi_e),\ee
or \be A_{\pm}^{\phi}={ (q_m+2n\phi_e)(\cos\theta\mp 1) \over (r^2+n^2)\sin^2\theta}.\ee
Notice that the first condition is important for satisfying the first law\footnote{This also was shown in \cite{pando} in a special case where $\phi_e=0$. } and the second is needed for obtaining the correct magnetic charge in the first law, which is also consistent with the path-integral conditions\footnote{Euclidean path-integral boundary conditions requires the regularity of the spatial components of the metric and gauge field, at the boundaries, i.e., radial infinity and the horizon. }. Notice also that the magnetic charge in the first law is different from the magnetic charge at radial infinity, $q_m$. The regularity of the gauge potential along the z-axis is equivalent to removing the whole z-axis from the enclosed volume and it carries a magnetic charge $-2n\phi_e$, as a result, we get $Q_m=Q_m^h$. This is the magnetic charge that contributes to thermodynamics and the first law.

Now let us calculate this magnetic charge directly from the nonsingular one-form $A$ after using Stock's theorem. This leads to \be Q_m=- {1 \over 4\pi} \oint A = - {1 \over 4\pi} \left(\int_{north-cap} A_{+} +\int_{south-cap} A_{-}\right)=q_m+2n\phi_e. \ee

Our conclusion is that the existence of the nut charge causes a difference between the magnetic charge at the horizon, which is relevant for thermodynamics, and the charge at radial infinity. The magnetic charge that contributes to the first law is also the one resulted from having finite gauge potential $A$, as required by Euclidean path integral. But what if we didn't impose the regularity of the gauge potential along the z-axis? In this case we will see that other electric and magnetic charges produce consistent thermodynamics, i.e., the first law, Gibbs-Duhem relation, and Smarr`s relation will be satisfied. These cases can be represented by a magnetic charge $ Q_m=q_m+\alpha n\phi_e$, as was discussed in \cite{adel_1,adel_2,Nutty_dyons_1,smarr}. Among the most important cases in this class is the case with $ Q_m=q_m+2 n\phi_e$, and $ Q_e= q_e$, as well as the self-dual case $ Q_m=q_m+n\phi_e$, and $ Q_e= q_e-n\phi_e$. In the following thermodynamic treatment and in our study of phase structure we are going to work with a one-form $A$ which is regular everywhere, this leads to the charges, $Q_m=q_m+2n\phi_e$, and $ Q_e= q_e$. 

\subsection{Thermodynamics}
Now we calculate thermal quantities, starting with temperature 
\begin{equation} \label{temp}
  T=\frac{ \left(1-\phi_e^2\right) r_0^2-\left( q_m + n\, \phi_e\right)^2}{4 \pi\,  r_0^3}\,+\,\frac{3\, r_0^2 \left(n^2+r_0^2\right)}{4 \pi\,  L^2\, r_0^3}.
\end{equation}
Again using the counter-terms method \cite{bala,surf99} one can calculate the on-shell gravitational action. It takes the form
\begin{equation}
   I=\beta \left( \frac{m}{2} + \frac{\left( q_m + n\, \phi_e\right)^2 - \phi_e^2 r_0^2}{2 r_0} - \frac{r_0^4 + 3 n^2 r_0^2}{2\,  L^2\, r_0} \right).
   \label{I_n}
\end{equation}
%             ************************
The mass of the solution is given by
\begin{equation}  \label{mbh}
    m=\frac{\left(q_e^2+q_m^2-n^2+r_0^2\right)}{2 r_0} + \frac{r_0^4+6 n^2 r_0^2-3 n^4}{2 L^2 r_0}.
\end{equation}
Recalling that the free energy is given by $I/\beta$, we substitute (\ref{mbh}) in (\ref{I_n}) and using (\ref{qe}) to get %the free energy of the ensemble
%$\beta=1/T$, we substitute (\ref{temp}) and (\ref{mbh}) in (\ref{I_n}) and using (\ref{qe}) to get the free energy of the ensemble
\be G =\frac{\left[\phi_e \left(n^2+r_0^2\right)+n q_m\right]^2}{4 r_0^3} - \frac{r_0^4 + 3 n^4}{4 L^2 r_0} +\frac{\left(n^2 - r_0^2\right) \left(2 \phi_e^2-1\right) + 3 q_m^2 + 4 n q_m \phi_e}{4 r_0}.  
 \label{fren}\ee

%           ***************************
It is important to see that the Euclidean path integral boundary conditions fix the boundary metric and the spatial component of the gauge potential, i.e. $g_{ij}$ and $A_i$, therefore, we have a mixed ensemble with the following partition function \be Z(\beta,n)=e^{-\beta\, G},\ee
or, $G=G(\beta,n,Q_m,\phi_e,P)$ (see discussion in \cite{adel_2}, section 2 on thermodynamic ensemble), where
\begin{equation}
dG= -\,S\,dT\, +\, \phi_n\,dn \,+\, \phi_m\, dQ_m\, -\,Q_e\, d\phi_e\, +\, V\, dP,
\end{equation}
and
\begin{eqnarray}
     && \left(\frac{\partial G}{\partial T}\right)_{n, Q_m,\phi_e,P} = -S, \;\;\;\;\;\;\;\;
        \left(\frac{\partial G}{\partial n}\right)_{T, Q_m,\phi_e,P} = \phi_n, \nonumber\\ \nonumber\\
       && \left(\frac{\partial G}{\partial P}\right)_{T,n, Q_m,\phi_e} =V, \;\;\;\;\;\;\;\;\;\;
      \left(\frac{\partial G}{\partial Q_m}\right)_{T,n,\phi_e,P}  = \phi_m,\nonumber\\  \nonumber\\ 
     && \;\;\;\; Q_e=-\left(\frac{\partial G}{\partial \phi_e}\right)_{T,n,Q_m,P}  = Q^{\infty}_e.
\end{eqnarray}
Calculating these quantities we get
\begin{multline} 
    \phi_n = \left(\frac{\partial G}{\partial n}\right)_{T, Q_m,\phi_e,P} \\ \\
    =\frac{n\,\left( Q_m - n\, \phi_e\right)^2\,+r_0^2\left(3\, n\, \phi_e^2 - n - 2\, Q_m\, \phi_e\right)}{2 r_0^3}\, +\, \frac{3\,n\,\left(r_0^2-n^2\right)}{2\,r_0\,L^2}, \;\;\;
      \label{chem2}
\end{multline} 
for the chemical potential of $n$. Also, the pressure $P$ and the volume $V$ are identical to the results of the neutral case. The magnetic potential is
\be \phi_m=\left(\frac{\partial G}{\partial Q_m}\right)_{T,n,\phi_e,P} =\frac{\left( Q_m - n\, \phi_e\right)}{r_0}, \ee 
while the internal energy is given by
\be
    U = M\,-\,n\, \phi_n-PV.
\ee

The above thermodynamic quantities should satisfy the following thermodynamic relations. First, the quantities satisfy the Gibbs-Duhem relation
\be
    G =  M\, -\,T\,S\, -\,n\, \phi_n \,-\, Q_e\, \phi_e.
\ee
All quantities satisfy Smarr's relation which can be put as
\be M=2\,T\,S +\,2\,n\, \phi_n \,+\,Q_e\, \phi_e\,+\, Q_m\,\phi_m\, -\,2\,P\, V.
\ee
More importantly, these quantities satisfy the first law 
\be
  dU=\,T\,dS\, +\, \phi_n\,dn \,+\, \phi_m\, dQ_m\, +\,\phi_e\, dQ_e\, -\, P\, dV. 
\ee

 %%%%%%%%%%%%%%%%%%%%%%%%%%%%%%%%%%%%%%%%%
 
\section{Dyonic Taub-NUT-AdS Phases}
In this section, we are going to study different phases that can emerge from dyonic Taub-NUT-AdS solutions. To keep our analysis tractable, it is important to constrain ourselves with two basic cases. The first case is the ensemble where $\phi_e=0$ while keeping $q_m$ and $n$ fixed, which we will call "canonical case". The second case is defined through $q_m=0$, while $n$ and $\phi_e$ are fixed, which we will call "mixed case".

\subsection{Canonical Case}
Now we begin our analysis by calculating the possible critical points of the equation of state, eqn.(\ref{temp}). For canonical case, since $\phi_e=0$, the equation of state reads
\be 
\label{barp}
\Bar{P} = \frac{\Bar{T}\,r_0^3 - r_0^2 + q_m^2} {r_0^2\left(r_0^2+n^2\right)},
\ee
where $\Bar{T}=4\pi T$ and $\Bar{P}=8\pi P$. %Following \cite{P-V} one can solve eqn.(\ref{barp}) together with
\begin{figure} [t!]
	\centering	
         \includegraphics[width=0.48 \textwidth] {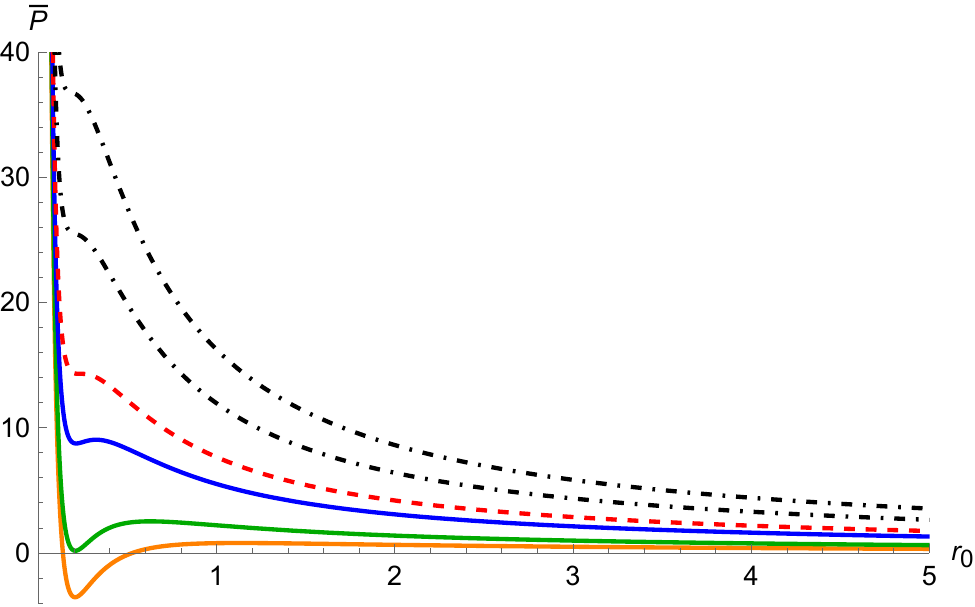}
         \includegraphics[width=0.48 \textwidth] {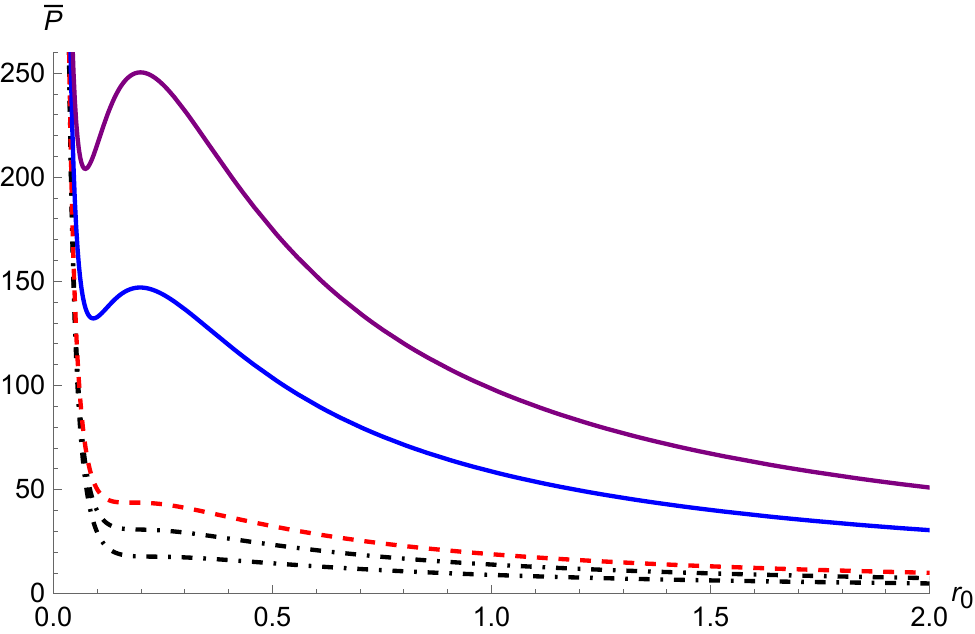}
         \includegraphics[scale=0.50] {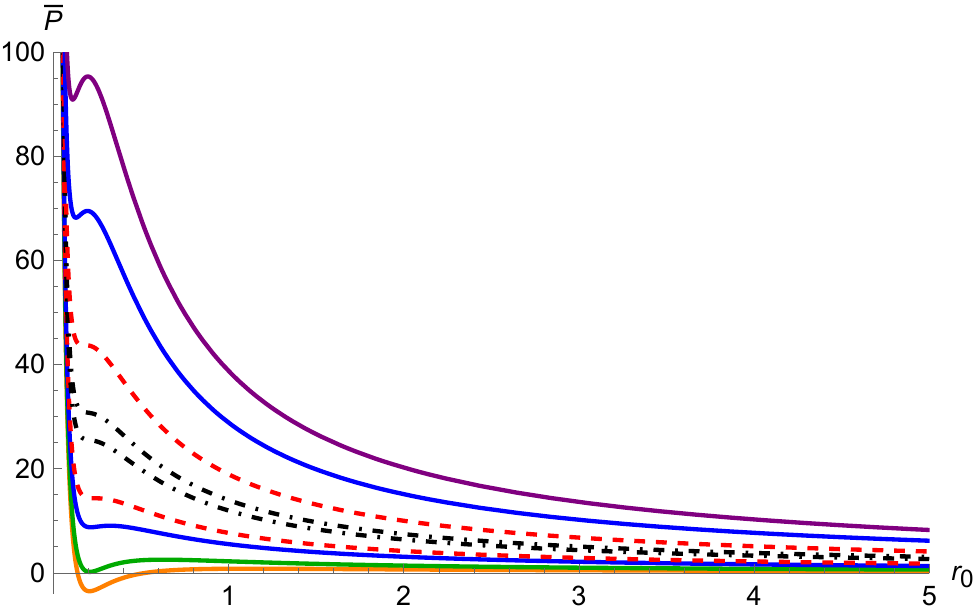}
         \caption{\footnotesize Behavior of canonical dyonic nut AdS solutions. In all panels, temperature decreases from top to bottom.  \textit{Top left}: $\Bar{P}-r_0$ diagram for the first critical point "a". The two upper dash-dotted isotherms correspond to continuous transition behavior (one-phase state) where $\Bar{T}>\Bar{T}_c^a$. The dashed line is the critical isotherm $\Bar{T}=\Bar{T}_c^a$. The lower isotherms correspond to the first-order transition where $\Bar{T}<\Bar{T}_c^a$. The first isotherm from below corresponds to $\Bar{T}<\Bar{T}_0$, followed by the $\Bar{T}_0$ isotherm, then $\Bar{T}_0<\Bar{T}<\Bar{T}_c^a$ isotherm. Isotherms from below corresponding to $\Bar{T}=.2 \Bar{T}_c^a, \Bar{T}_0, 0.75 \Bar{T}_c^a, \Bar{T}_c^a, 1.5 \Bar{T}_c^a$, and $2 \Bar{T}_c^a$. \textit{Top right}: Diagram represents the second critical point "b". Here the one-phase isotherms correspond to $\Bar{T}<\Bar{T}_c^b$, these are the two dash-dotted lines, while the two-phases correspond to $\Bar{T}>\Bar{T}_c^b$, which are the two solid lines. The dashed line is the critical isotherm $\Bar{T}=\Bar{T}_c^b$. Isotherms from below: $\Bar{T}=0.5 \Bar{T}_c^b, 0.75 \Bar{T}_c^b, \Bar{T}_c^b, 3 \Bar{T}_c^b$, and $5 \Bar{T}_c^b$. \textit{Bottom}: Diagram shows isotherms for both critical points.  Isotherms from below: $\Bar{T}=.2 \Bar{T}_c^a, \Bar{T}_0, 0.75 \Bar{T}_c^a, \Bar{T}_c^a, 1.5 \Bar{T}_c^a, 0.75 \Bar{T}_c^b, \Bar{T}_c^b, 1.5 \Bar{T}_c^b$, and $2 \Bar{T}_c^b$. In calculations we considered $n = 0.2$ and $q_m = 0.12$.}
 	\label {p-v-can}
\end{figure}
Solving eqn.(\ref{barp}) together with $\frac{\del{\Bar{P}}}{\del r_0} = 0$ and $\frac{\del^2{\Bar{P}}} {\del r_0^2} = 0$, one gets the following four solutions
\begin{equation}
  \begin{split}
    \Bar{P}_c^{(1,2)} & = \frac{1}{n^4} \left(6 q_m^2-n^2+2 \sqrt{3} \sqrt{q_m^2 \left(3 q_m^2-n^2\right)}\right) , \\ 
    \Bar{T}_c^{(1,2)} & = \pm \frac{1}{3 n^4}
    \left[16 \left(3 q_m^2-\sqrt{3} \sqrt{q_m^2 \left(3 q_m^2-n^2\right)}\right)^{3/2} \right. \\ 
   & \;\;\;\;\;\;\;\;\;\;\;\;\;\;\;\; \left. + \left(8 n^2-96 q_m^2\right) \sqrt{3 q_m^2-\sqrt{3} \sqrt{q_m^2 \left(3 q_m^2-n^2\right)}}\;\right], \\ 
    r_{0(c)}^{(1,2)} & = \mp{\sqrt{3 q_m^2-\sqrt{3} \sqrt{q_m^2 \left(3 q_m^2-n^2\right)}}},
  \end{split}
  \label{cpcana}
\end{equation}

and
\begin{equation}
   \begin{split}
    \Bar{P}_c^{(3,4)} & = \frac{1}{n^4} \left(6 q_m^2-n^2-2 \sqrt{3} \sqrt{q_m^2 \left(3 q_m^2-n^2\right)}\right) , \\ 
    \Bar{T}_c^{(3,4)} & = \pm \frac{1}{3 n^4} \left[16 \left(3 q_m^2+\sqrt{3} \sqrt{q_m^2 \left(3 q_m^2-n^2\right)}\right)^{3/2} \right.\\
    & \;\;\;\;\;\;\;\;\;\;\;\;\;\;\;\; \left. + \left(8 n^2-96 q_m^2\right) \sqrt{3 q_m^2+\sqrt{3} \sqrt{q_m^2 \left(3 q_m^2-n^2\right)}}\;\right],    \\ 
     r_{0(c)}^{(3,4)} & = \mp \sqrt{3 q_m^2+\sqrt{3} \sqrt{q_m^2 \left(3 q_m^2-n^2\right)}}.
   \end{split}
   \label{cpcanb}
\end{equation}

Only two of the above four solutions are physical, i.e., those with positive radii and temperatures. A special case occurs when $q_m=\frac{n}{\sqrt{3}}$, in which case the square root 
\begin{equation}
    \sqrt{q_m^2 \left(3 q_m^2-n^2\right)}
\end{equation}
vanishes and we get two duplicated solutions, from which only one is physical, i.e., we get one merged critical point.

The existence of two critical points renders this analysis qualitatively different from the one that uses a different thermodynamic approach in \cite{nutty_dyons_2}. Note that these expressions reduce to the known solutions in the case $n=0$. As an example let us choose $n=0.2$ and $q_m=0.12$, then one gets two critical points
\begin{eqnarray}
    &  \Bar{P}_c^a = 14.303 \,\, , \,\,\,\,\, \Bar{T}_c^a = 8.942 \,\, , \,\,\,\,\, r_{0(c)}^a = 0.2344, \\
    &  \Bar{P}_c^b = 43.697 \,\, , \,\,\,\,\, \Bar{T}_c^b = 20.662 \,\, , \,\,\,\,\, r_{0(c)}^b = 0.1773 .
\end{eqnarray}
The corresponding AdS radii for these critical points are 
\be L_c^a = 0.458 \,\, , \,\,\, \& \,\,\, L_c^b = 0.262 . \ee

The corresponding $\Bar{P}-r_0$ diagram is displayed in Fig. \ref{p-v-can}. Evidently, for $n \neq 0$ there are two critical points $a$ and $b$. When $\Bar{T} < \Bar{T}_c^a$ the behavior resembles Van der Waals fluid, see the left top panel of Fig. \ref{p-v-can}. In this region, a first-order phase transition between the small and the large radii occurs. Such transition is governed by Maxwell's equal area law, which guarantees two-phase coexistence when the areas above and below the isobar drawn through the $\Bar{P}-r_0$ curve are equal. Also, there is a temperature $\Bar{T}_0$, 
\be
\Bar{T}_0 = \frac{r_0^2 - q_m^2}{r_0^3}   \label{T0can}, \ee
below which the pressure is negative for some $r_0$, which was discussed in \cite{P-V}, where this region marked unphysical. We are going to do the same here and will not consider this region since in any case it corresponds to de Sitter (dS) solutions rather than anti-de Sitter AdS ones.

As $\Bar{T}$ increases above $\Bar{T}_c^a$, there exists a region of continuous transition where only one radius exists, which continues to happen as long as we have $\Bar{T} < \Bar{T}_c^b$, and till the second critical isotherm is reached. As $\Bar{T}$ exceeds $\Bar{T}_c^b$, as in the right top panel, the system imitates the Van der Waals fluid again, meaning it retains the first-order phase transition between the small and the large radii. We can see the behavior before and after both critical points together in the bottom panel of the figure.

\subsubsection{Phase Structure}
\begin{figure} [h!]
	\centering
		\includegraphics[width=0.48 \textwidth]{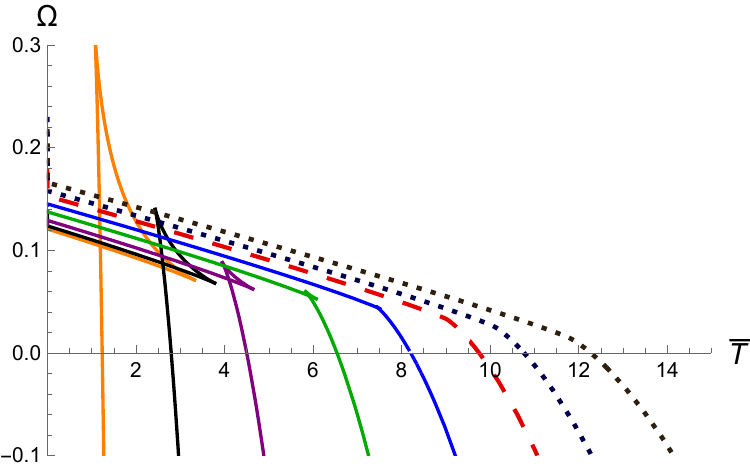}
           \includegraphics[width=0.48 \textwidth]
          {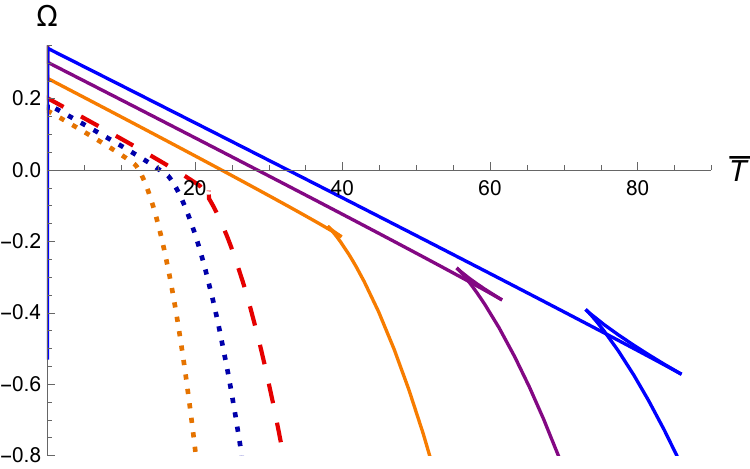}
	\caption{\footnotesize Free energy as a function of temperature for the canonical dyonic AdS solutions plotted at different values of Pressure for $n = .2$ and $q_m = .12$. Pressure increases from left to right in both panels. \textit{Left}: First-order transition for $\Bar{P}<\Bar{P}_c^a$ is characterized by the swallowtail behavior represented by the solid lines. The dotted lines represent the continuous transition for $\Bar{P}>\Bar{P}_c^a$. The dashed line corresponding to $\Bar{P}=\Bar{P}_c^a$. Curves from left to right are plotted for $\Bar{P}=.05 \Bar{P}_c^a, 0.1 \Bar{P}_c^a, 0.25 \Bar{P}_c^a, 0.5 \Bar{P}_c^a, \Bar{P}_c^a, 1.2 \Bar{P}_c^a$ and $1.5 \Bar{P}_c^a$. \textit{Right}: The small-large radii first-order transition occurs for $\Bar{P}>\Bar{P}_c^b$, solid lines. The two dotted lines for the continuous transition are now for $\Bar{P}<\Bar{P}_c^b$. The dashed line corresponding to $\Bar{P}=\Bar{P}_c^b$. Curves from left to right are plotted for $\Bar{P}=0.5 \Bar{P}_c^b, 0.75 \Bar{P}_c^b, \Bar{P}_c^b, 2 \Bar{P}_c^b$, $3 \Bar{P}_c^b$ and $4 \Bar{P}_c^b$.}
 	\label {canEpc}
\end{figure}

To investigate the phase structure of Taub-Nut solutions we must study its free energy. For canonical ensemble the free energy is given from eqn.(\ref{fren}) to be
\begin{equation}
     \Omega =\frac{r_0^2\left(r_0^2 - n^2\right) + 3 r_0^2 q_m^2 + n^2 q_m^2}{4 r_0^3} - \frac{r_0^4 + 3 n^4}{4 L^2 r_0}
 \label{can_E}
\end{equation}
In Fig. \ref{canEpc} we plotted the free energy as a function of temperature at different values of pressure $\Bar{P}$. 

\begin{figure} % [h!]
	\centering
		\includegraphics[width=0.5 \textwidth] {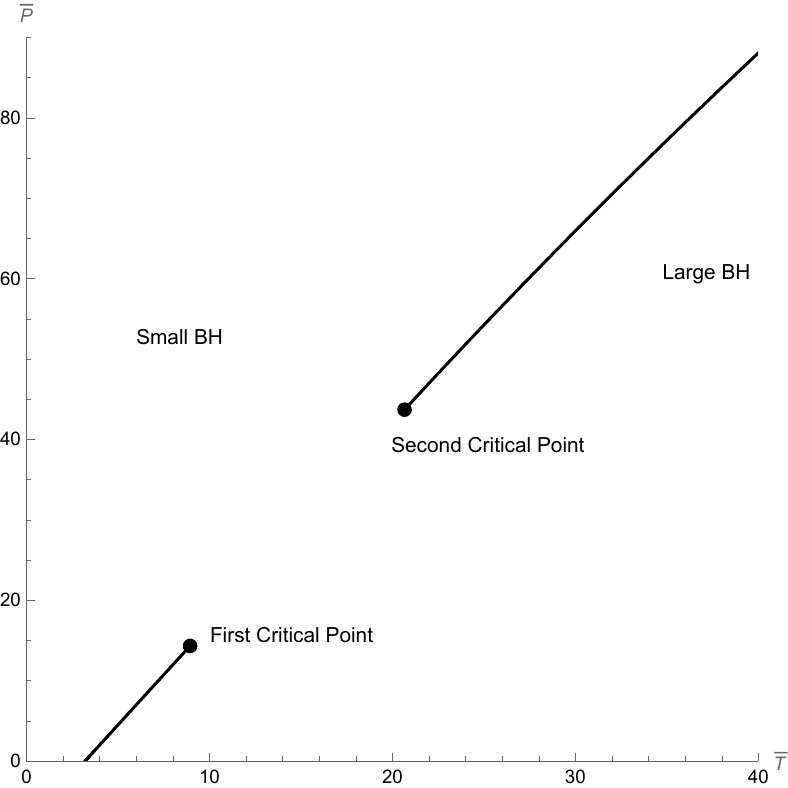}
	\caption{\footnotesize Phase diagram of the canonical Taub-Nut AdS solutions calculated for $n = 0.2$ and $q_m=0.12$ showing first-order phase transition for $\Bar{P} < \Bar{P}_c^a$ and $\Bar{P} > \Bar{P}_c^b$. For $\Bar{P}_c^a < \Bar{P} < \Bar{P}_c^b$, a continuous transition occurs. The two critical points $(\Bar{T}_c^a, \Bar{P}_c^a)$ and $(\Bar{T}_c^b, \Bar{P}_c^b)$ are declared by dots.} 
 	\label {canphas}
\end{figure}

The free energy, as seen from Fig. \ref{canEpc}, is characterized by a swallowtail shape when $\Bar{P}<\Bar{P}_c^a$, as shown in the left panel, and for $\Bar{P}>\Bar{P}_c^b$, in the right panel. This behavior shows the existence of two critical points, not one as in the usual Van der Waals fluids, or the RN-AdS cases studied in \cite{clifford_cata,P-V}, with a continuous-phase transition region trapped between them. %Certain values of pressure and temperature show a 
The swallowtail behavior, as shown by Fig. \ref{canEpc}, takes place in the temperature intervals $\left[\Bar{T}_1^a, \Bar{T}_2^a\right]$ for $\Bar{P}<\Bar{P}_c^a$ and $\left[\Bar{T}_1^b, \Bar{T}_2^b\right]$ for $\Bar{P}>\Bar{P}_c^b$. Increasing pressure results in diminishing the temperature range $\left[\Bar{T}_1^a, \Bar{T}_2^a\right]$, as seen by the left panel, in contrast, increasing the temperature range $\left[\Bar{T}_1^b, \Bar{T}_2^b\right]$, as seen by the right panel, while $\Bar{T}_1$ and $\Bar{T}_2$ coincide at $\Bar{P} = \Bar{P}_c$. Inside each of these temperature ranges there exists a transition temperature, $\Bar{T}_{tr} \in \left[\Bar{T}_1, \Bar{T}_2\right]$, at which the free energies of the small and the large horizon radii match. In the $\Bar{P} - \Bar{T}$ plane, the transition temperatures form a curve through which phase transitions occur between solutions with small and large horizon radii, but solutions with large radii possess smaller free energy, therefore it is the stable phase. This is depicted in Fig. \ref{canphas}.

%Accordingly, this is a first-order phase transition. 
The line of phase transition can be found using Maxwell’s equal area law, or by finding the points in the $\Bar{P} - \Bar{T}$ plane for which the free energy and temperature coincide for small $(r_0 = r_s)$ and large $(r_0 = r_l)$ radii. Following the last method, the radii of the small and large solutions at which the transition occurs are found to be
\begin{equation} \label{rslc}
    r_s=\frac{1}{2} \left(-y + \sqrt{y^2 + 4 x}\right) ,  \hspace{.6 in} r_l=\frac{x}{r_s},
\end{equation}
where $y$ is given by the relation
\begin{equation}  \label{yrc}
    y=\sqrt{\frac{-1}{q_m^2} \left[\Bar{P} x^3 - \left(\Bar{P} n^2 + 1\right) x^2 - 3\; q_m\; x\right]},
\end{equation}
and $x$ is one of the roots of the equation
\begin{equation}
    \Bar{P}^2 x^4 - \left(n^2 \Bar{P} + 1\right) \Bar{P} x^3 + 3 \left(n^2 \Bar{P} + 1\right) q_m^2 x - 9 q_m^4 = 0 .
\end{equation}
The transition temperature $\Bar{T}_{tr}$ is then obtained by substituting $r_s$, or $r_l$, in the temperature relation 
\begin{equation} \label{Tcan}
    \Bar{T}_{tr} = \frac{1}{r_s^3} \left[r_s^2\left(r_s^2+n^2\right) \Bar{P} + r_s^2 - q_m^2 \right]
\end{equation}
which is obtained from (\ref{barp}).

As the pressure comes to its first critical value $\Bar{P}=\Bar{P}_c^a$, the swallowtail behavior vanishes, keeping a kink on the free energy curve. Increasing the pressure further, $\Bar{P} > \Bar{P}_c^a$, the transition between the small and large solutions is now continuous, and the free energy curve becomes a monotonic curve. The kink at the critical pressure in the free energy curve characterizes a point at the end of the first transition curve in the $\Bar{P} - \Bar{T}$ plane, where a continuous transition begins to occur, as seen in Fig. \ref{canphas}. This continuous transition occupies a region in the $\Bar{P} - \Bar{T}$ plane where  $\Bar{P}_c^a < \Bar{P} < \Bar{P}_c^b$. At the point $\Bar{P} = \Bar{P}_c^b$, the kink appears again on the free energy curve and the two boundaries of the temperature interval $\left[\Bar{T}_1^b, \Bar{T}_2^b\right]$ are now coinciding. Increasing the pressure to exceed $\Bar{P}_c^b$ renders the system to have a first-order transition again. As the right panel of Fig. \ref{canEpc} indicates, the interval $\left[\Bar{T}_1^b, \Bar{T}_2^b\right]$ is now widened as $\Bar{P}$ increased. A new transition curve in the $\Bar{P}-T$ plane begins at the point $(\Bar{P}_c^b, \Bar{T}_c^b)$ and continues endlessly, Fig. \ref{canphas}. The first-order phase transition occurs again along this curve from the small to the large radii.
%%%%%%%%%%%%%%%%%%%%%%%%%%%%%%%%%%%%%%%%%%%%%%%%%%%%%%%%

\subsubsection{Special Case: Merged Critical Points}
\begin{figure}  [t!]
	\centering
		\includegraphics[width=0.48 \textwidth] {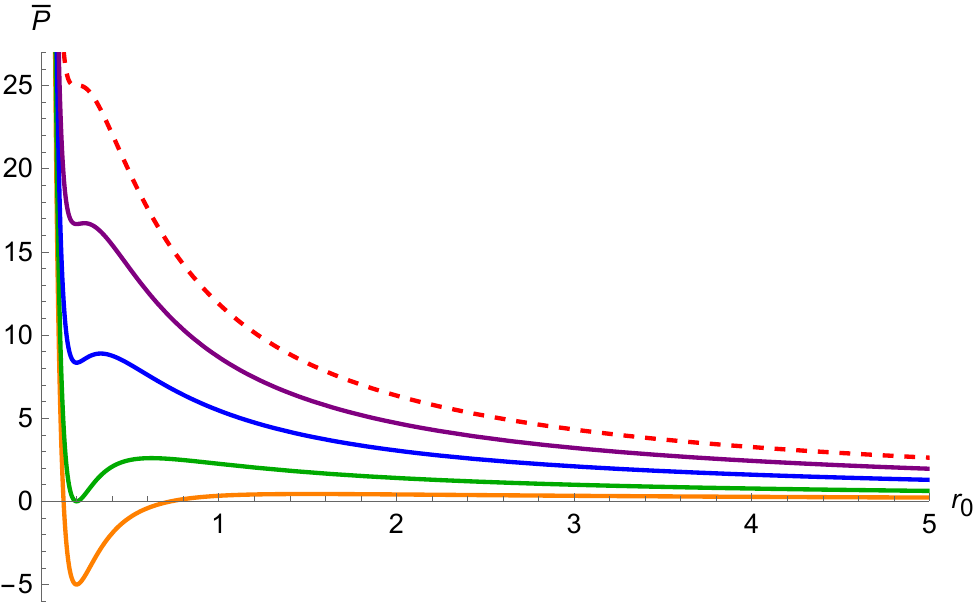}
        \includegraphics[width=0.48 \textwidth] {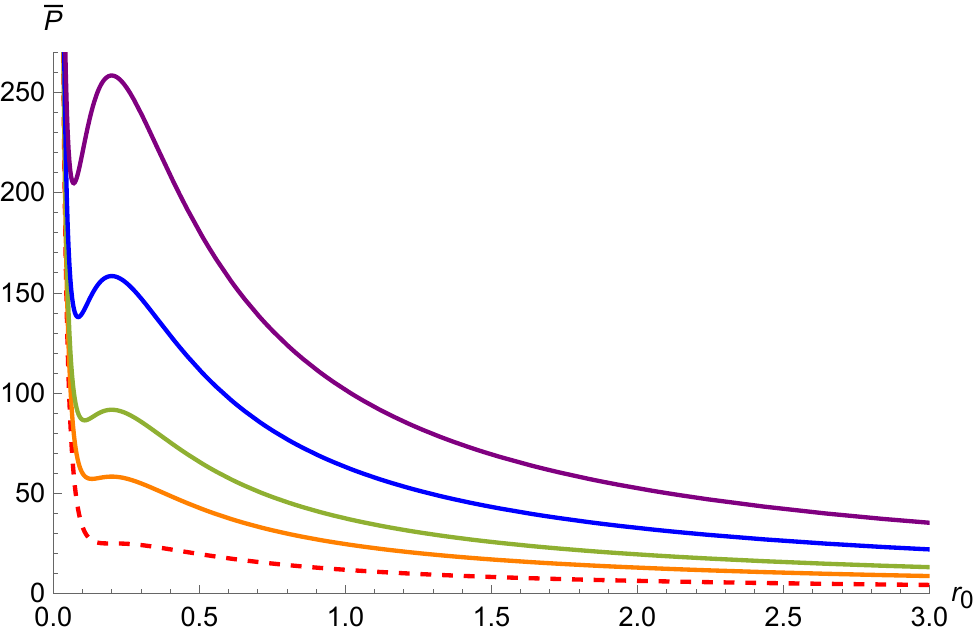}
        \includegraphics[width=0.50 \textwidth] {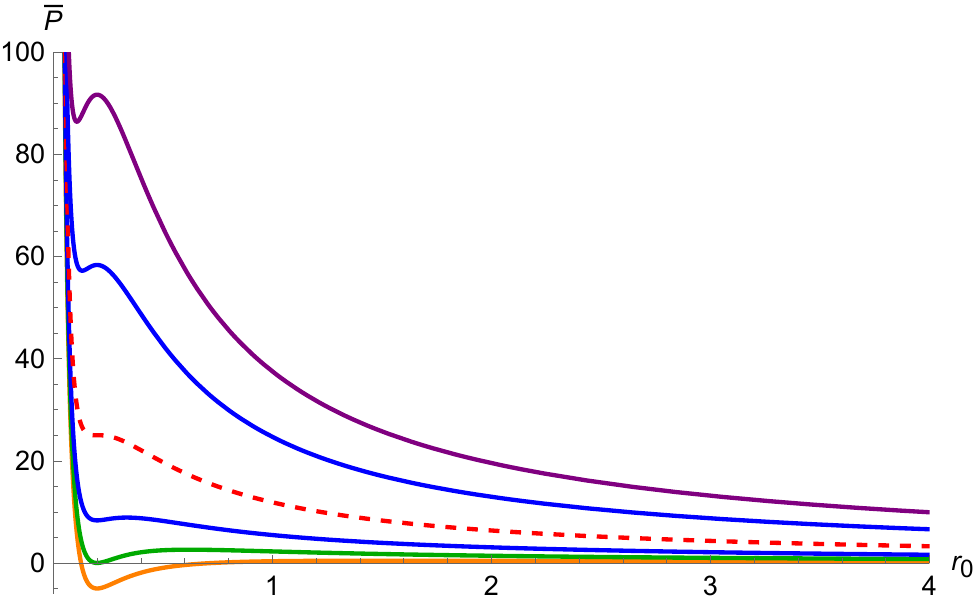}   
        \caption{\footnotesize Behavior of canonical dyonic nut AdS solutions when $q_m=\frac{n}{\sqrt{3}}$. Isotherms decrease the temperature from top to bottom in all panels. \textit{Top left}: Diagram for $\Bar{T} \leq \Bar{T}_c$. The dashed line represents a critical isotherm. Isotherms from below corresponding to $\Bar{T}=.1 \Bar{T}_c, \Bar{T}_0, 0.5 \Bar{T}_c, 0.75 \Bar{T}_c$, and $\Bar{T}_c$. \textit{Top right}: Diagram for $\Bar{T} \geq \Bar{T}_c$. Isotherms from below: $\Bar{T}=\Bar{T}_c, 2 \Bar{T}_c, 3 \Bar{T}_c, 5 \Bar{T}_c$, and $8 \Bar{T}_c$. \textit{Bottom}: General behavior for merged critical point. Isotherms from below: $\Bar{T}=.1 \Bar{T}_c, \Bar{T}_0, 0.5 \Bar{T}_c, \Bar{T}_c, 2 \Bar{T}_c$, and $3 \Bar{T}_c$.} 
 	\label {mergTcan}
\end{figure}

When $q_m = \frac{n}{\sqrt{3}}$, A new phenomenon occurs. The two critical points merge into one. All isotherms before and after the critical isotherm resemble isotherms of RN-AdS solutions. The first-order phase transition occurs for any temperature except $\Bar{T}_c$. This is clear in Fig. \ref{mergTcan}, where we displayed the $\Bar{P}-r_0$ diagram for $n = .2$.

\begin{figure} %[h!]
	\centering
		\includegraphics[width=0.50 \textwidth] {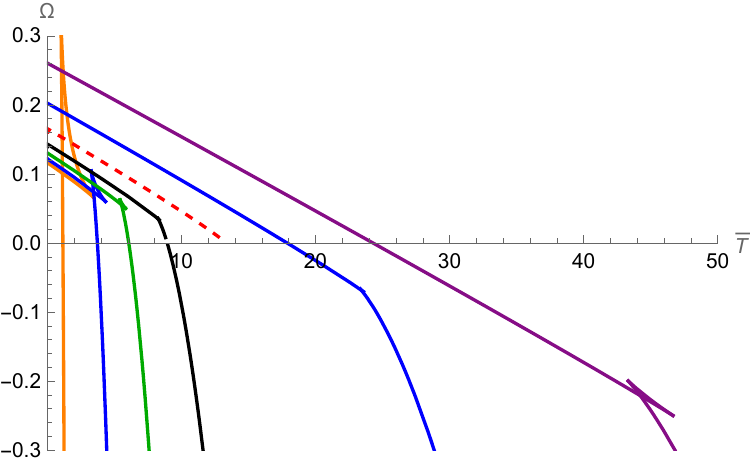}
	\caption{\footnotesize Free energy as a function of temperature for different values of Pressure for the $n = 0.2$. The dashed curve is plotted at $\Bar{P}_c$. Curves from left to right are plotted for $\Bar{P}=.01 \Bar{P}_c, 0.1 \Bar{P}_c, 0.25 \Bar{P}_c, 0.5 \Bar{P}_c$, $\Bar{P}_c$, 2 $\Bar{P}_c$ and 4 $\Bar{P}_c$.}
 	\label {mergEcans}
\end{figure}

The free energy curves confirm this phenomenon. In Fig. \ref{mergEcans} we plot the free energy curves for the whole pressure range. The figure shows that the swallowtail behavior characterizing the first-order phase transition always exists no matter $\Bar{P}<\Bar{P}_c$ or $\Bar{P}>\Bar{P}_c$. However, relations give imaginary results for most $\Bar{T}$ range when $\Bar{P}=\Bar{P}_c$. 
\begin{figure}  [h!]
	\centering
		\includegraphics[width=0.5 \textwidth] {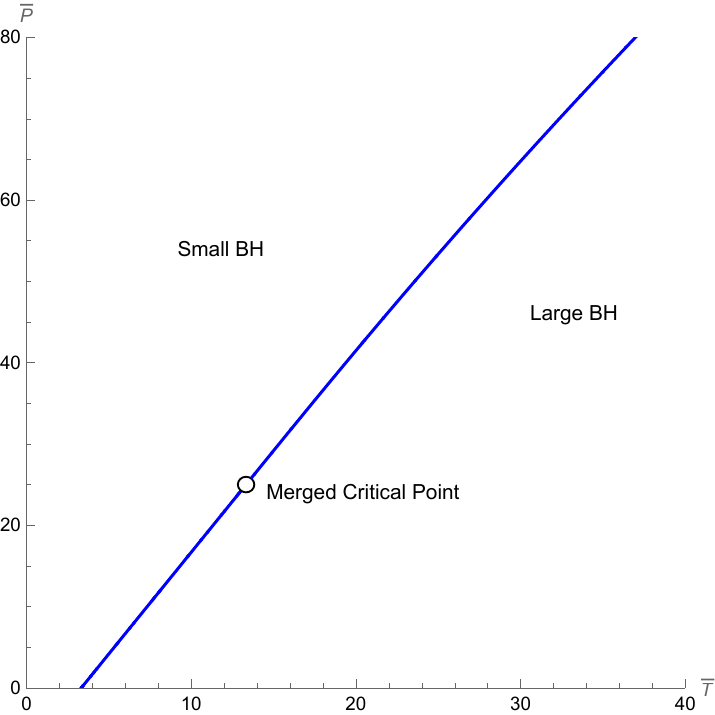}
	\caption{\footnotesize Phase diagram of the canonical dyonic Taub-Nut AdS solutions for merged critical points for $n = 0.1$ showing endlessly first-order phase transition for all $\Bar{P}$ except for $\Bar{P}=\Bar{P}_c$. The critical point $(\Bar{T}_c, \Bar{P}_c)$ is declared by the circle.}
 	\label {mergphase}
\end{figure}

The corresponding phase diagram is plotted in Fig. \ref{mergphase} for $n=0.1$. The circle on the curve represents the position of the critical point. The graph shows that except for the critical point, The first-order phase transition is the only phase transition that occurs.

%%%%%%%%%%%%%%%%%%%%%%%%%%%%%%%%%%%%%%%%%%%%%%%

\subsection{Mixed Case}
For the mixed ensemble case, we have $q_m=0$. The equation of state reduces to
\be \Bar{P} = \frac{\Bar{T}\,r_0^3 + n^2\, \phi_e^2 + (\phi_e^2 - 1) r_0^2}{r_0^2\left(r_0^2+n^2\right)}  \label{eosm} \ee
Critical points are then obtained by solving this relation together with $\frac{\del{\Bar{P}}}{\del r_0} = 0$ and $\frac{\del^2{\Bar{P}}} {\del r_0^2} = 0$. In doing so we get the four solutions
\begin{equation}
  \begin{split}
    \Bar{P}_c^{(1,2)} & = 
 \left(\frac{1 - \phi_e^2}{n^4}\right)  \left(\frac{6 n^2 \phi_e^2}{\phi_e^2-1} + \frac{2 \sqrt{3} \sqrt{n^4 \left(4 \phi_e^4-\phi_e^2\right)}}{\phi_e^2-1} \right)
    -\frac{\left(1 - 13\phi_e^2\right)}{n^2} , \\ 
   \Bar{T}_c^{(1,2)} & =\pm \frac{1}{3 n^4}  \left[16 \left(1 - \phi_e^2\right) \left(-\frac{\sqrt{3} \sqrt{n^4 \left(4 \phi_e^4-\phi_e^2\right)}}{\phi_e^2-1}-\frac{3 n^2 \phi_e^2}{\phi_e^2-1}\right)^{3/2} \right. \\ 
   & \;\;\;\;\;\;\;\;\;\;\;\;\;\; \left. + n^2 \left(8-104\phi_e^2\right) \sqrt{-\frac{\sqrt{3} \sqrt{n^4 \left(4 \phi_e^4-\phi_e^2\right)}}{\phi_e^2-1}-\frac{3 n^2 \phi_e^2}{\phi_e^2-1}} \;\right], \\
     r_{0(c)}^{(1,2)} & =\pm \sqrt{-\frac{\sqrt{3} \sqrt{n^4 \left(4 \phi_e^4-\phi_e^2\right)}}{\phi_e^2-1}-\frac{3 n^2 \phi_e^2}{\phi_e^2-1}}.
  \end{split}
  \label{pm12}
\end{equation}
\\
and
\begin{equation}
  \begin{split}
    \Bar{P}_c^{(3,4)} & = 
 \left(\frac{1 - \phi_e^2}{n^4}\right)  \left(\frac{6 n^2 \phi_e^2}{\phi_e^2-1} - \frac{2 \sqrt{3} \sqrt{n^4 \left(4 \phi_e^4-\phi_e^2\right)}}{\phi_e^2-1} \right)
    -\frac{\left(1 - 13\phi_e^2\right)}{n^2} , \\ 
    \Bar{T}_c^{(3,4)} & =\pm \frac{1}{3 n^4}  \left[16 \left(1 - \phi_e^2\right) \left(\frac{\sqrt{3} \sqrt{n^4 \left(4 \phi_e^4-\phi_e^2\right)}}{\phi_e^2-1}-\frac{3 n^2 \phi_e^2}{\phi_e^2-1}\right)^{3/2} \right. \\
  & \;\;\;\;\;\;\;\;\;\;\;\;\;\; \left. + n^2 \left(8-104\phi_e^2\right) \sqrt{\frac{\sqrt{3} \sqrt{n^4 \left(4 \phi_e^4-\phi_e^2\right)}}{\phi_e^2-1}-\frac{3 n^2 \phi_e^2}{\phi_e^2-1}} \;\right], \\
     r_{0(c)}^{(3,4)} & =\pm \sqrt{\frac{\sqrt{3} \sqrt{n^4 \left(4 \phi_e^4-\phi_e^2\right)}}{\phi_e^2-1}-\frac{3 n^2 \phi_e^2}{\phi_e^2-1}}.  \\ \\
  \end{split}
  \label{pm34}
\end{equation}

Again, only two of these four solutions are physical, i.e. those with positive radii and temperatures. Another property that specifies the critical points of the mixed ensemble is the fact that the number of the critical points of this ensemble depends on the value of $\phi_e$. To see this let's analyze the square roots that exist in the solutions. One main square root is
\begin{equation} \label{cond1}
    n^2 |\phi_e| \sqrt{ \left(4 \phi_e^2-1\right)}.
\end{equation}
Due to this root, physical critical points exist only if $\phi_e^2 \ge 1/4$. For $\phi_e = \pm 1/2$, the square root vanishes and the two solutions coincide. In this case, the two critical points merge irrespective of the value of the nut charge $n$. 

Another square root exists in the critical temperatures and radii, namely
\begin{equation}  \label{cond2a}
    \sqrt{\pm \frac{\sqrt{3} \sqrt{n^4 \left(4 \phi_e^4-\phi_e^2\right)}}{\phi_e^2-1}-\frac{3 n^2 \phi_e^2}{\phi_e^2-1}}.
\end{equation}
The existence of a real solution for this root depends on the sign of the denominator. If the denominator is negative, in which case $\abs{\phi_e} < 1$, terms will reverse signs. The square root in this case is correctly rewritten as
\begin{equation}   \label{cond2b}
    \sqrt{\mp \frac{\sqrt{3} \sqrt{n^4 \left(4 \phi_e^4-\phi_e^2\right)}}{1-\phi_e^2}+\frac{3 n^2 \phi_e^2}{1-\phi_e^2}}.
\end{equation}
Accordingly, if the sign of the first term is positive, the root is always real. If, on the other hand, the sign of the first term is negative, a condition is put on the terms, i.e.,
\begin{equation}    \label{cond2c}
    3 n^2 \phi_e^2 \ge \sqrt{3} \sqrt{n^4 \left(4 \phi_e^4-\phi_e^2\right)},
\end{equation}
to ensure a real solution. Analyzing this we come to the condition $\abs{\phi_e} \le 1$, which is consistent with our starting condition since we have to exclude the value of $\abs{\phi_e} = 1$ as it blows up our solutions. This guarantees the existence of the other critical point.

On the other hand, a positive denominator in eqn.(\ref{cond2a}) demands the condition $\abs{\phi_e} > 1$ on the potential. In this case, one solution is always imaginary, while the other is real only if
\begin{equation}    \label{cond2d}
     \sqrt{3} \sqrt{n^4 \left(4 \phi_e^4-\phi_e^2\right)} \ge 3 n^2 \phi_e^2,
\end{equation}
Which is verified only if $\abs{\phi_e} \ge 1$. Since  this is a consistent solution (we again exclude the case $\abs{\phi_e} = 1$), we then guarantee the existence of one critical point.

A final case is characterized by $\abs{\phi_e} = 1$, which lead an equation of state
\be \Bar{P} = \frac{\Bar{T}\,r_0^3 + n^2}{r_0^2\left(r_0^2+n^2\right)}.  \label{eosms} \ee
Solving for critical points we get the two solutions
\begin{eqnarray} \label{cpspecial}
    \Bar{P}_c^{(1)} = \frac{12}{n^2}, \;\;\;\;\;\;\; \Bar{T}_c^{(1)} = \frac{-16 \sqrt{2}}{n}, \;\;\;\;\;\;\; r_{0(c)}^{(1)} = \frac{-n}{\sqrt{2}},   \\
    \Bar{P}_c^{(2)} = \frac{12}{n^2}, \;\;\;\;\;\;\; \Bar{T}_c^{(2)} = \frac{16 \sqrt{2}}{n}, \;\;\;\;\;\;\; r_{0(c)}^{(2)} = \frac{n}{\sqrt{2}}. 
\end{eqnarray}    
Evidently, only one of these solutions is physical, i.e., we have only one critical point.

Summarizing our results for mixed ensemble, we have the following cases:
\begin{enumerate}
    \item $1/2 < \abs{\phi_e} < 1$ : two critical points exist,
    \item $\abs{\phi_e} = 1/2$ : A one duplicated, merged, critical point exist,
    \item $\abs{\phi_e} > 1$ : A single critical point exist,
    \item $\abs{\phi_e} = 1$ : A single critical point exist.  
\end{enumerate}

%%%%%%%%%%%%%%%%%%%%%%%%%%%%%%%%%%%%%%%%%%%%%%%%%%%%%%%%%%%%%%%%%%%%%%

\subsubsection{Case I: $1/2 < \phi_e < 1$}
In this case, the system has two critical points, setting $n=0.1$ and $\phi_e=0.53$, we get
\begin{eqnarray}
    &  \Bar{P}_c^a = 32.08 \,\, , \,\,\,\,\, \Bar{T}_c^a = 10.89 \,\, , \,\,\,\,\, r_{0(c)}^a = .127 , \\
    &  \Bar{P}_c^b = 161.18 \,\, , \,\,\,\,\, \Bar{\Bar{T}_c^b} = 36.55 \,\, , \,\,\,\,\, r_{0(c)}^b = .085 .
\end{eqnarray}
The corresponding AdS scales for these critical points are 
\be L_c^a = 0.306 \, , \,\,\,\; \& \,\,\,\,\; L_c^b = 0.136 . \ee
\begin{figure} [t!] 
	\centering
		\includegraphics[width=0.45 \textwidth] {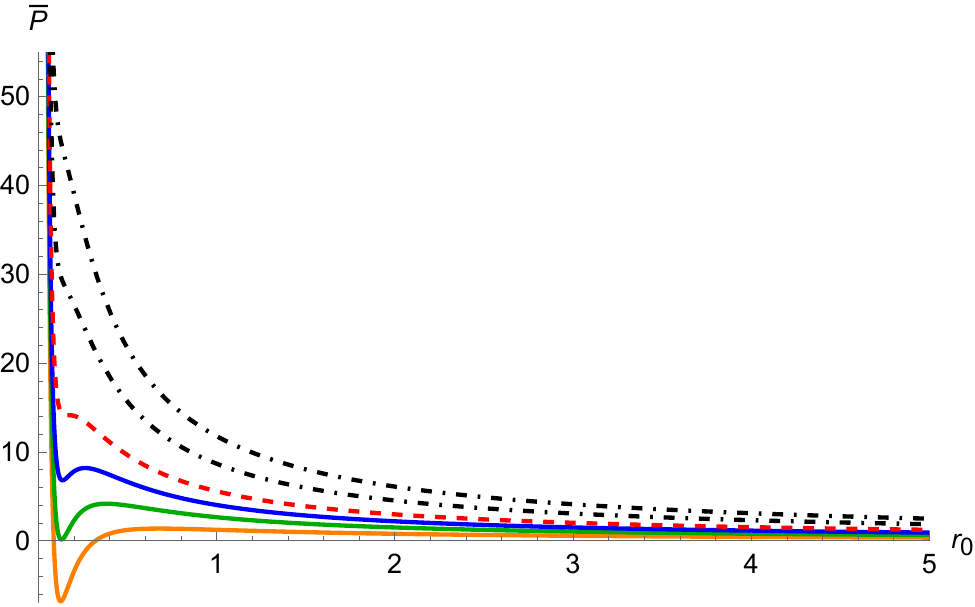}
		\includegraphics[width=0.45 \textwidth] {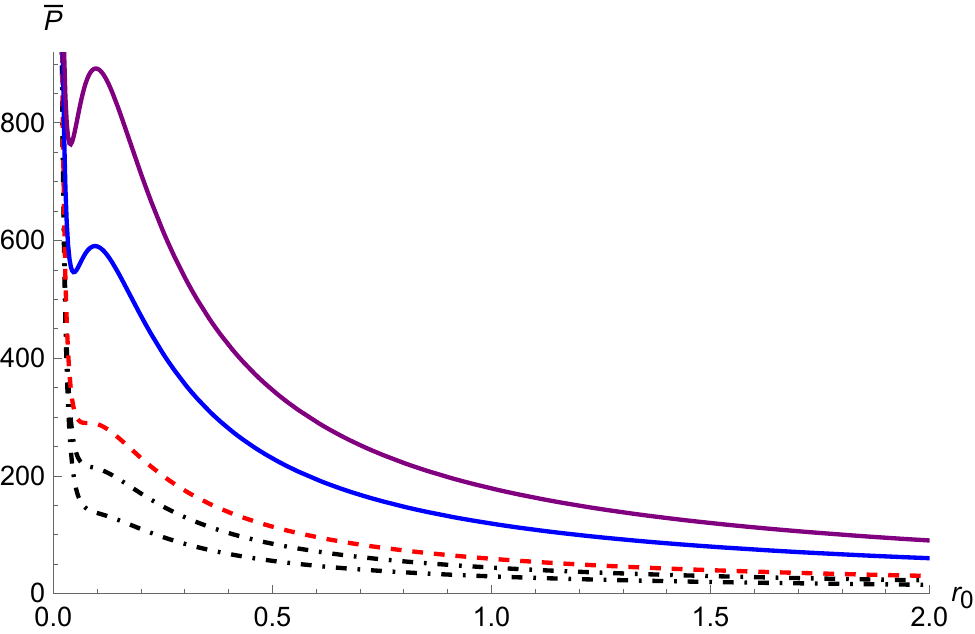}
            \includegraphics[width=0.55 \textwidth]{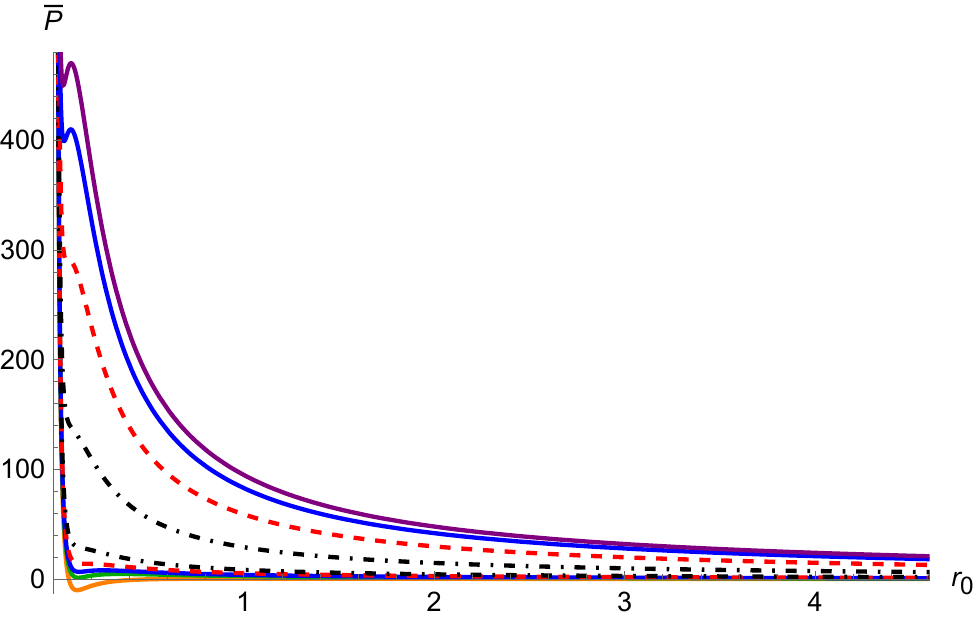}
    	\caption{\footnotesize Behavior of the mixed dyonic nut AdS solutions for $1/2 < \phi_e < 1$. In all panels, temperature decreases from top to bottom.  \textit{Top left}: $\Bar{P}-r_0$ diagram for the first critical point "a". The two upper dash-dot isotherms correspond to continuous transition behavior where $\Bar{T}>\Bar{T}_c^a$. The dashed line is the critical isotherm $\Bar{T}=\Bar{T}_c^a$. The lower isotherms correspond to the first-order transition where $\Bar{T}<\Bar{T}_c^a$. The first isotherm from below corresponds to $\Bar{T}<\Bar{T}_0$, followed by the $\Bar{T}_0$ isotherm, then $\Bar{T}_0<\Bar{T}<\Bar{T}_c^a$ isotherm. Isotherms from below corresponding to $\Bar{T}=.3 \Bar{T}_c^a, \Bar{T}_0, .75 \Bar{T}_c^a, \Bar{T}_c^a, 1.5 \Bar{T}_c^a$, and $2 \Bar{T}_c^a$. \textit{Top right}: Diagram for the second critical point "b". For this critical point, the one-phase isotherms correspond to $\Bar{T}<\Bar{T}_c^b$, which are represented by the two dash-dotted lines in the bottom, while the two-phases correspond to $\Bar{T}>\Bar{T}_c^b$, which are represented by the two solid lines on top. The dashed line is the critical isotherm $\Bar{T}=\Bar{T}_c^b$. Isotherms from below: $\Bar{T}=.5 \Bar{T}_c^b, .75 \Bar{T}_c^b, \Bar{T}_c^b, 2 \Bar{T}_c^b$, and $3 \Bar{T}_c^b$. \textit{Bottom}: Isotherms for both critical points.  Isotherms from below: $\Bar{T}=.2 \Bar{T}_c^a, \Bar{T}_0, .75 \Bar{T}_c^a, \Bar{T}_c^a, 1.5 \Bar{T}_c^a, .5 \Bar{T}_c^b, \Bar{T}_c^b, 1.4 \Bar{T}_c^b$, and $1.6 \Bar{T}_c^b$. In calculations we considered $n = 0.1$ and $\phi_e = 0.6$.}
         \label {mixT}
\end{figure}

In Fig. \ref{mixT} we display the corresponding $\Bar{P}-r_0$ for such solutions. Isotherms around each critical point are plotted separately in the top panel, while in the bottom panel we represent isotherms for both critical points together. We can see that for $\Bar{P} < \Bar{P}_c^a$ and $\Bar{P} > \Bar{P}_c^b$, isotherms resemble the Van der Waals fluids. In these regions, a first-order phase transition takes place between solutions with small and large radii which is controlled by Maxwell's equal area law. While for $\Bar{P}_c^a \le \Bar{P} \le \Bar{P}_c^b$, a continuous phase transition takes place, i.e., we have a single-phase state, and a single solution exists at any temperature. The temperature $\Bar{T}_0$ below which the pressure is negative for some $r_0$ is now given by 

\be
T_0 = \frac{1-\phi_e^2}{r_0}-\frac{n^2 \phi_e^2}{r_0^3}   \label{T0mix} \ee
%%%%%%%%%%%%%%%%%%%%%%%%%%%%%%%%%%%%%%%%%%%%%%%%%%%%%%%%%%

\subsubsection{Phase Structure}
The free energy for a mixed ensemble of Taub-Nut AdS solutions is given by replacing $q_m=0$ in eq.(\ref{fren})
\begin{equation}
     \Omega = \frac{\phi_e^2 \left(n^2+r_0^2\right)^2 + r_0^2 \left(n^2 - r_0^2\right) \left(2 \phi_e^2-1\right)}{4 r_0^3} - \frac{r_0^4 + 3 n^4}{4 L^2 r_0} 
 \label{freemix}
\end{equation}

\begin{figure} [h!]
	\centering
		\includegraphics[width=0.48 \textwidth] {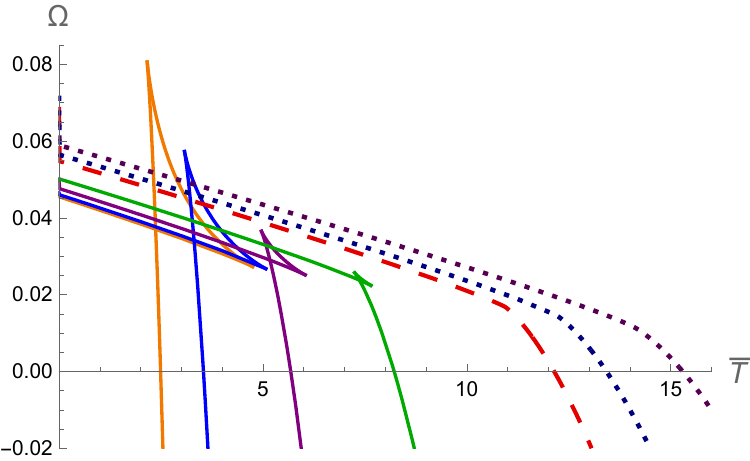}
             \includegraphics[width=0.48 \textwidth] {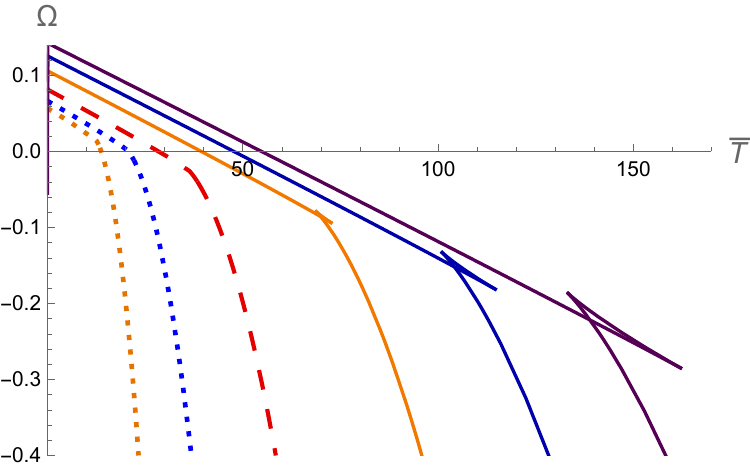}
	\caption{\footnotesize Free energy as a function of temperature for the mixed dyonic AdS solutions plotted at different values of pressure for $n = .1$ and $\phi = .53$. Pressure increased from left to right for both panels. \textit{Left}: Swallowtail behavior characterizes the first-order transition for $\Bar{P}<\Bar{P}_c^a$ represented by the solid lines. The dotted lines represent the continuous-phase transition for $\Bar{P}>\Bar{P}_c^a$ cases. The dashed line represents the $\Bar{P}=\Bar{P}_c^a$ case. Curves from left to right are plotted for $\Bar{P}=.05 \Bar{P}_c^a, 0.1 \Bar{P}_c^a, 0.25 \Bar{P}_c^a, 0.5 \Bar{P}_c^a, \Bar{P}_c^a, 1.2 \Bar{P}_c^a$ and $1.5 \Bar{P}_c^a$. \textit{Right}: The small to large radii first-order transition which occurs for $\Bar{P}>\Bar{P}_c^b$ is represented by the solid lines. The two dotted lines for the continuous-phase transition represent the $\Bar{P}<\Bar{P}_c^b$ cases. The dashed line represents the $\Bar{P}=\Bar{P}_c^b$ case. Curves from left to right are plotted for $\Bar{P}=.25 \Bar{P}_c^b, 0.5 \Bar{P}_c^b, \Bar{P}_c^b, 2 \Bar{P}_c^b, 3 \Bar{P}_c^b,$ and $4 \Bar{P}_c^b$.}
 	\label {mixEpc1}
\end{figure}

In Fig. \ref{mixEpc1} we present the free energy as a function of temperature for different values of the pressure $\Bar{P}$. The swallowtail behavior appears clearly for $\Bar{P}<\Bar{P}_c^a$, in the left panel, and $\Bar{P}>\Bar{P}_c^b$, in the right panel. Similar to the canonical ensemble case, a first-order phase transition occurs in these two regions which is controlled by Maxwell's equal area law. Following the same method as before, the radii of the small and large solutions at which the transition occurs are

\begin{figure} [h!]
	\centering
            \includegraphics[width=0.5 \textwidth] {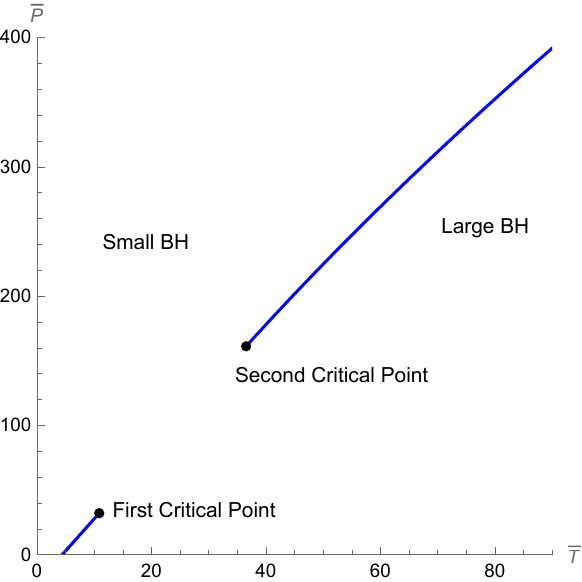}
	\caption{\footnotesize Phase diagram of the mixed dyonic Taub-Nut AdS solutions for $1/2<\phi_e<1$ showing two critical points with a  continuous transition occurring in between. Here we set $n = 0.1$ and $\phi_e=0.53$. The two critical points $\left(\Bar{T}_c^a, \Bar{P}_c^a\right)$ and $\left(\Bar{T}_c^b, \Bar{P}_c^b\right)$ are declared by the dots.} 
 	\label {mixphs}
\end{figure}

\begin{equation} \label{rsl}
    r_s=\frac{1}{2} \left(-y + \sqrt{y^2 + 4 x}\right) ,  \;\;\;\;\;\;\;\;\; r_l=\frac{x}{r_s} ,
\end{equation}
where $y$ is given by the relation
\begin{equation}  \label{yr}
    y=\sqrt{\frac{-1}{n^2 \phi_e^2} \left\{\Bar{P} x^3 - 3 \left[\Bar{P} n^2 + \left(1 - \phi_e^2\right)\right] x^2 - 3\; n^2 \phi_e^2\; x\right\}} ,
\end{equation}
and $x$ is one of the roots of the equation
\begin{equation}
    \Bar{P}^2 x^4 - \left[n^2 \Bar{P} + \left(1 - \phi_e^2\right)\right] \Bar{P} x^3 + 3 n^2 \phi_e^2 \left[n^2 \Bar{P} + \left(1 - \phi_e^2\right)\right] x - 9 n^2 \phi_e^4 = 0 .
\end{equation}
The transition temperature $\Bar{T}_{tr}$ is then obtained by substituting $r_s$, or $r_l$, in the temperature relation 
\begin{equation} \label{Tmix}
    \Bar{T}_{tr} = \frac{1}{r_s^3} \left[r_s^2\left(r_s^2+n^2\right) \Bar{P} + r_s^2 \left(1 - \phi_e^2\right) - n^2 \phi_e^2\right],
\end{equation}
which is obtained from (\ref{eosm}).

Starting with $\Bar{P}<\Bar{P}_c^a$, as we increase the pressure to reach a point where $\Bar{P}=\Bar{P}_c^a$, the transition between the small and large radii becomes continuous announcing a continuous phase transition in this region. If we increased the pressure more, where $\Bar{P}>\Bar{P}_c^b$, an abrupt transition between the small and large radii will take place, and the system goes through a first-order phase transition once again. The phase diagram is shown in Fig. \ref{mixphs}. 

%%%%%%%%%%%%%%%%%%%%%%%%%%%%%%%%%%%%%%%%%%
\subsubsection{Case II: $\phi_e = 1/2$}
When $\phi_e = 1/2$, the two solutions merged into one. For example, for $n=0.1$ we get
\begin{equation}
    \Bar{P}_c = 75 \,\, , \,\,\,\,\, \Bar{T}_c = 20 \,\, , \,\,\,\,\, r_{0(c)} = 0.1 .
\end{equation}
The corresponding AdS scale for this critical point is
\be L_c = 0.2 . \ee 

\begin{figure} [hbt!]
	\centering		
         \includegraphics[width=0.45 \textwidth] {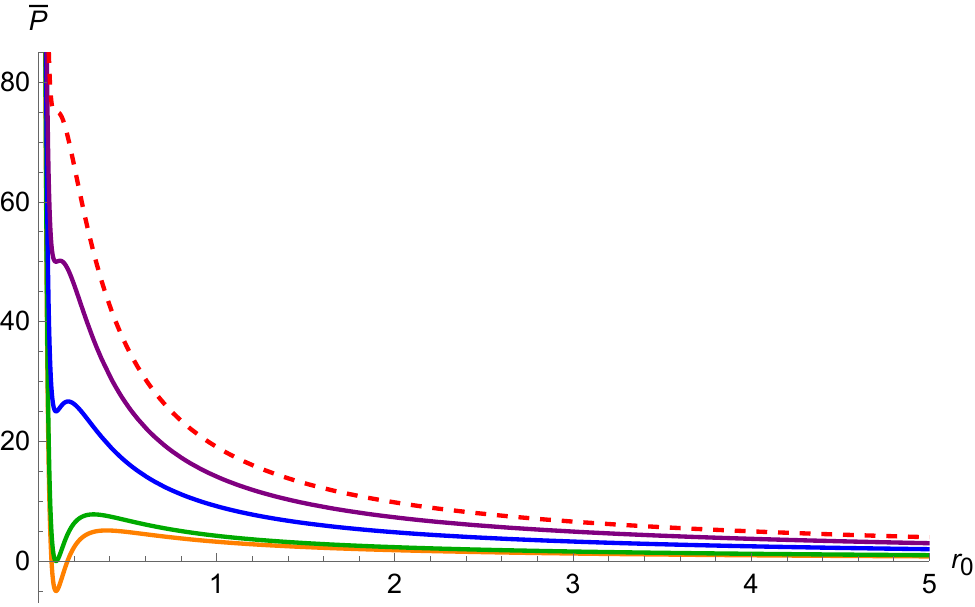}
        \includegraphics[width=0.45 \textwidth] {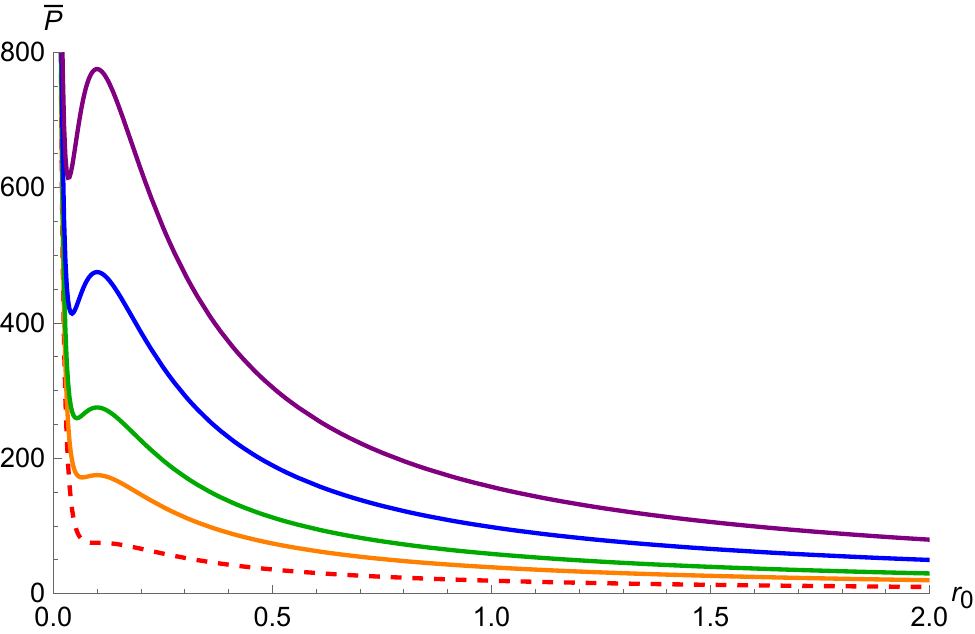}
        \includegraphics[width=0.55 \textwidth] {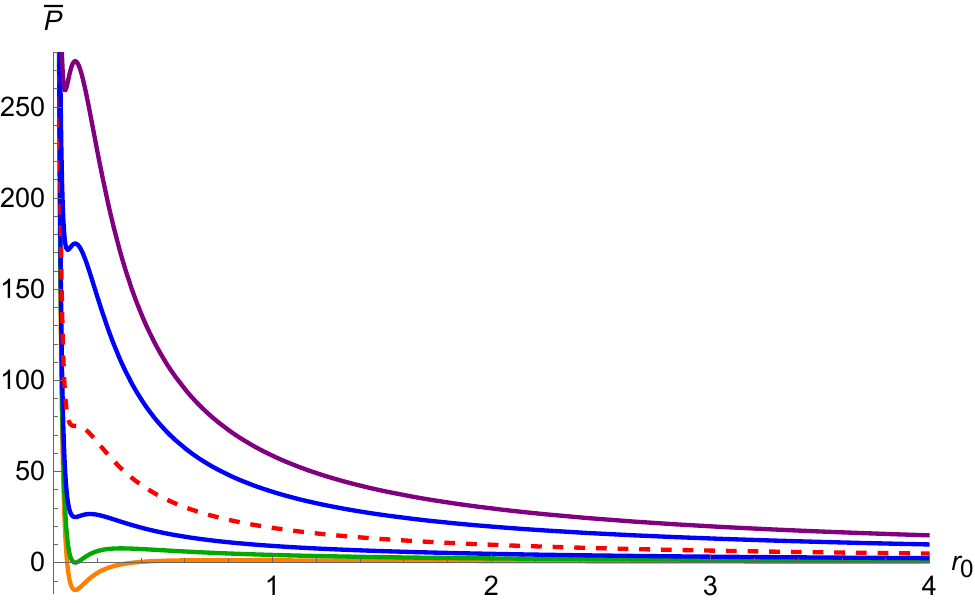}
    	\caption{\footnotesize Behavior of mixed dyonic nut solutions for merged critical point, $\phi_e = 1/2$. Isotherms decrease the temperature from top to bottom for each panel. The dashed line corresponds to a critical isotherm. \textit{Top left}: $\Bar{P}-r_0$ diagram for $\Bar{T} \le \Bar{T}_c$. Isotherms from below corresponding to $\Bar{T}=.2 \Bar{T}_c, \Bar{T}_0, 0.5 \Bar{T}_c, 0.75 \Bar{T}_c$, and $\Bar{T}_c$. \textit{Top right}: Diagram for $\Bar{T} \ge \Bar{T}_c$. Isotherms from below: $\Bar{T}= \Bar{T}_c, 2 \Bar{T}_c, 3 \Bar{T}_c, 5 \Bar{T}_c$, and $8 \Bar{T}_c$. \textit{Bottom}: Isotherms mimicking Van der Waals fluids for both $\Bar{T} < \Bar{T}_c$ and $\Bar{T} > \Bar{T}_c$. Isotherms from below are plotted for $\Bar{T}=.1 \Bar{T}_c, \Bar{T}_0, 0.5 \Bar{T}_c, \Bar{T}_c, 2 \Bar{T}_c$, and $3 \Bar{T}_c$. In calculations, $n = 0.1$.} 
 	\label {p-v-mix-merge}
\end{figure}
%[scale=0.55]
%, where the region of the continuous transition in between the critical points disappears as they are merged

The region between the two critical points in the previous case disappears as the two critical points become one. As a consequence, the continuous-phase transition exists only at the critical point, as indicated in Fig. \ref{p-v-mix-merge}. The three solutions corresponding to the Van der Waals fluid occur after the critical isotherm as well as before it. Only at the critical isotherm we can see the continuous phase transition.

\begin{figure} [t!]
	\centering
		\includegraphics[width=0.45 \textwidth] {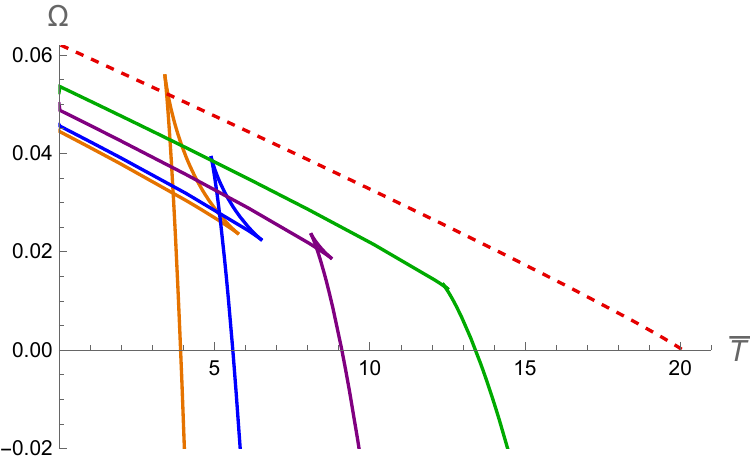}
		\includegraphics[width=0.45 \textwidth] {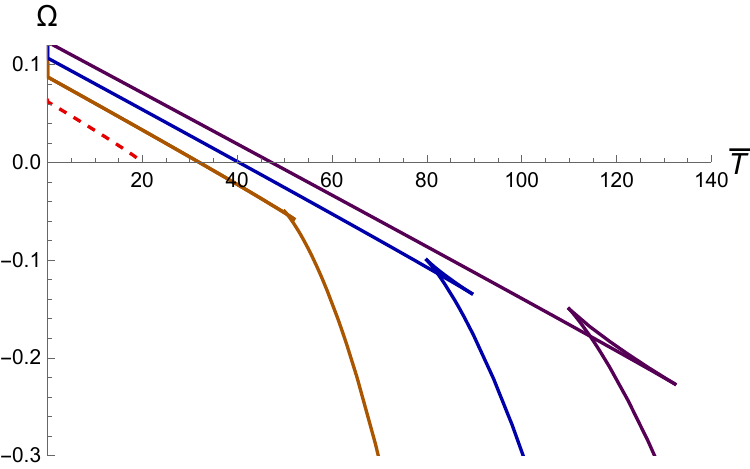}
	\caption{\footnotesize Free energy as a function of temperature for the mixed ensemble at $n = 0.1$ and $\phi_e = 1/2$. The dashed lines corresponding to $\Bar{P}=\Bar{P}_c$. \textit{Left}: First-order transition recognized by the swallowtail behavior for $\Bar{P}<\Bar{P}_c$. Curves from left to right represent $\Bar{P}=.04 \Bar{P}_c, 0.08 \Bar{P}_c, 0.2 \Bar{P}_c, 0.5 \Bar{P}_c$ and $\Bar{P}_c$. \textit{Right}: Swallowtail behavior returns back for all $\Bar{P}>\Bar{P}_c$. Curves from left to right represent $\Bar{P}=\Bar{P}_c, 2 \Bar{P}_c, 3 \Bar{P}_c, 5 \Bar{P}_c$ and $7 \Bar{P}_c$.}
 	\label {mixEmrg}
\end{figure}

In the free energy diagram, we note the swallowtail phenomenon throughout the whole range of pressure, except for $\Bar{P}=\Bar{P}_c$, see Fig. \ref{mixEmrg}.
\begin{figure} [h!]
	\centering
		\includegraphics[width=0.5 \textwidth] {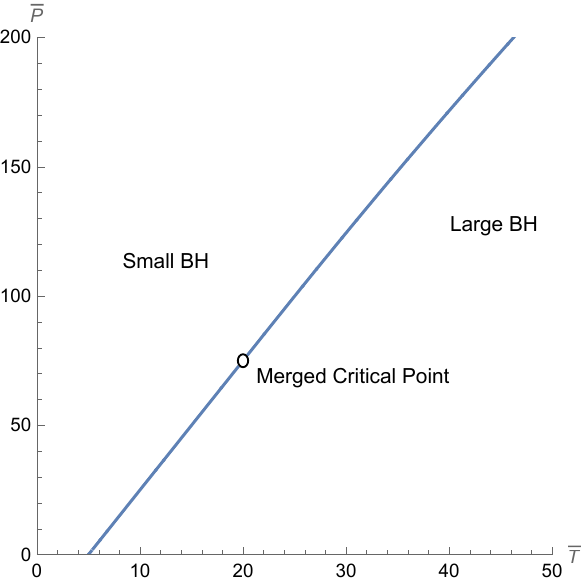}
	\caption{\footnotesize Phase diagram of the mixed dyonic Taub-Nut AdS solutions at $\phi_e=1/2$ for $n = 0.1$ showing a first-order phase transition continues endlessly for all values of $\Bar{P}$ except for $\Bar{P}=\Bar{P}_c$. The critical point $(\Bar{T}_c, \Bar{P}_c)$ is declared by the circle.} 
 	\label {mixphsmrg}
\end{figure}
The phase diagram is plotted in Fig. \ref{mixphsmrg}, where we can see that except for the critical point, the first-order phase transition from the small to the large radii continues to take place.

%%%%%%%%%%%%%%%%%%%%%%%%%%%%%%%%%%%%%%%%%%%%%%%%%%%%%

\subsubsection{Case III: $\phi_e > 1$}
\begin{figure} [t!]
	\centering
		\includegraphics[width=0.45 \textwidth] {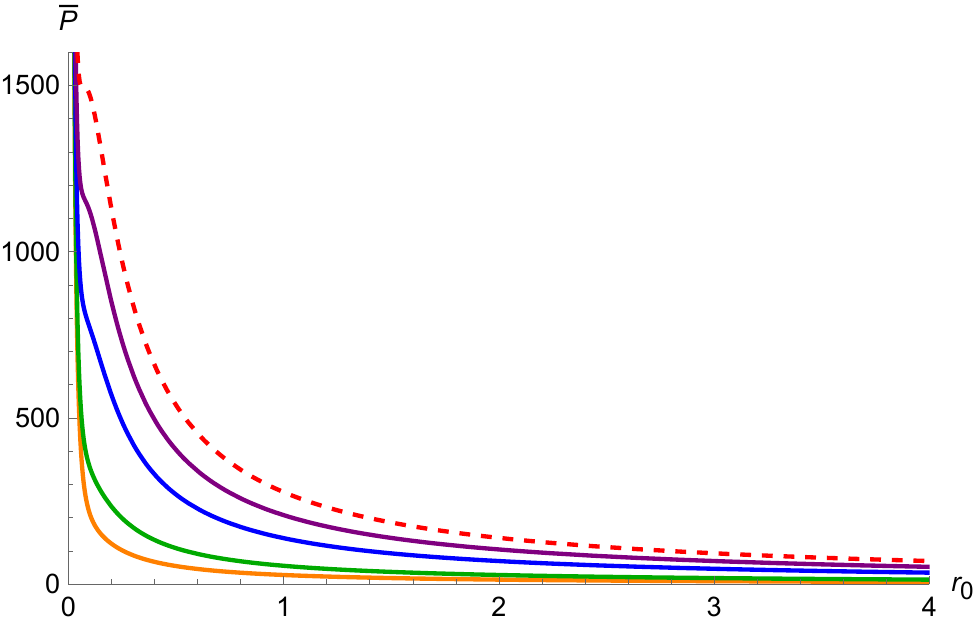}
             \includegraphics[width=0.45 \textwidth] {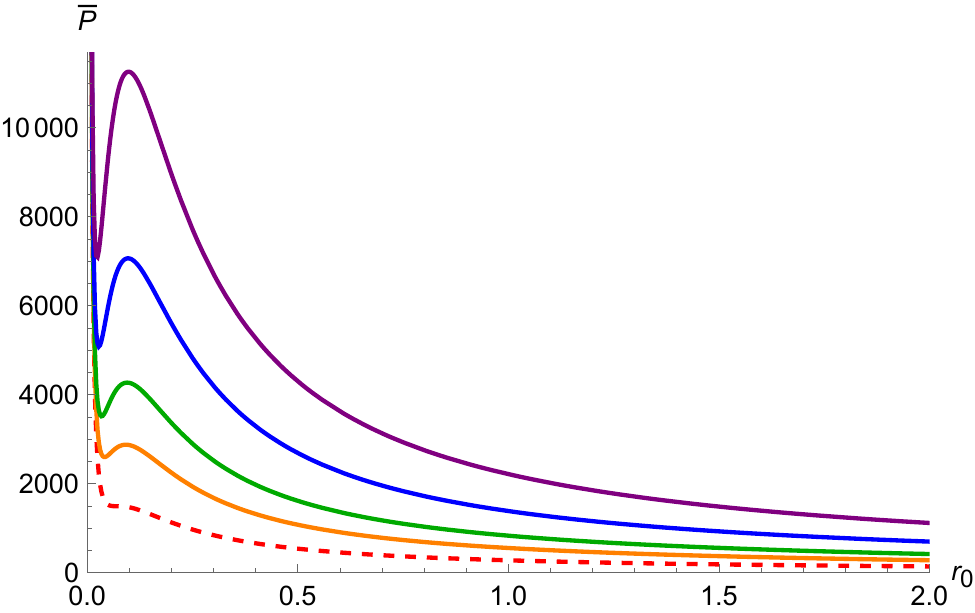}
             \includegraphics[width=0.55 \textwidth] {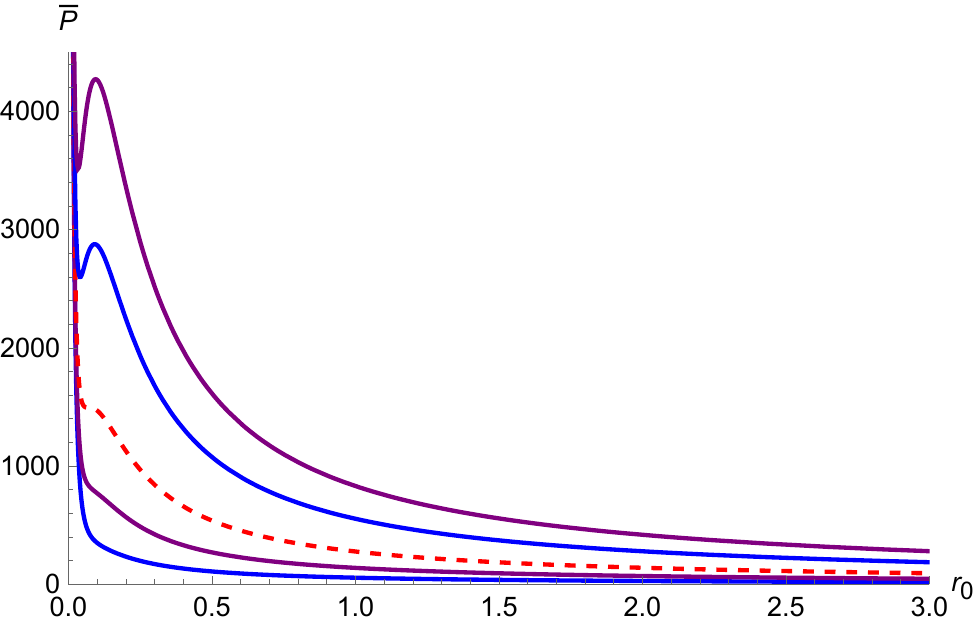}
    	\caption{\footnotesize Mixed dyonic nut solutions showing single critical point behavior for $\phi_e > 1$. Isotherms decrease the temperature from top to bottom for each panel. In calculations $n = 0.1$ and $\phi_e = 1.1$. The dashed line corresponds to a critical isotherm. \textit{Top left}: Continuous transition for $\Bar{T} \le \Bar{T}_c$. Isotherms from below corresponding to $\Bar{T}=.1 \Bar{T}_c, 0.2 \Bar{T}_c, 0.5 \Bar{T}_c, 0.75 \Bar{T}_c$, and $\Bar{T}_c$. \textit{Top right}: First-order phase transition behavior for $\Bar{T} > \Bar{T}_c$. Isotherms from below: $\Bar{T}= \Bar{T}_c, 2 \Bar{T}_c, 3 \Bar{T}_c, 5 \Bar{T}_c$, and $8 \Bar{T}_c$. \textit{Bottom}: Single critical point behavior for $\phi_e > 1$. Isotherms mimicking Van der Waals fluids for $\Bar{T} > \Bar{T}_c$ only. Isotherms from below: $\Bar{T}=.2 \Bar{T}_c, 0.5 \Bar{T}_c, \Bar{T}_c, 2 \Bar{T}_c$, and $3 \Bar{T}_c$.}
 	\label {mix-unq}
\end{figure}

When $\phi_e > 1$, a single critical point exists showing a very new behavior, namely, the continuous phase transition region in the $P-T$ diagram is close to the origin, in contrast with what happens in Reissner-Nordstrom-AdS solutions and Van der Waals fluids, i.e., the continuous phase transition happens only for low enough pressures and temperatures! Isotherms now resemble Van der Waals fluid behavior only for $\Bar{T} > \Bar{T}_c$. We can see in this region the existence of the three solutions with different radii (two of them are thermodynamically stable), see the top right panel of Fig. \ref{mix-unq}.
A first-order phase transition then occurs from the small to the large radii for these temperatures, which is controlled by Maxwell's equal area law. The region where $\Bar{T} \le \Bar{T}_c$, on the other hand, is a region of continuous transition, top left panel of Fig. \ref{mix-unq}. Notice that we do not have a temperature $\Bar{T}_0$ below which the pressure is negative for some $r_0$ as in previous cases. The value of $\Bar{T}_0$ becomes negative, see eqn.(\ref{T0mix}), i.e., is unphysical. 

\begin{figure} [t!]
	\centering
		\includegraphics[width=0.5 \textwidth] {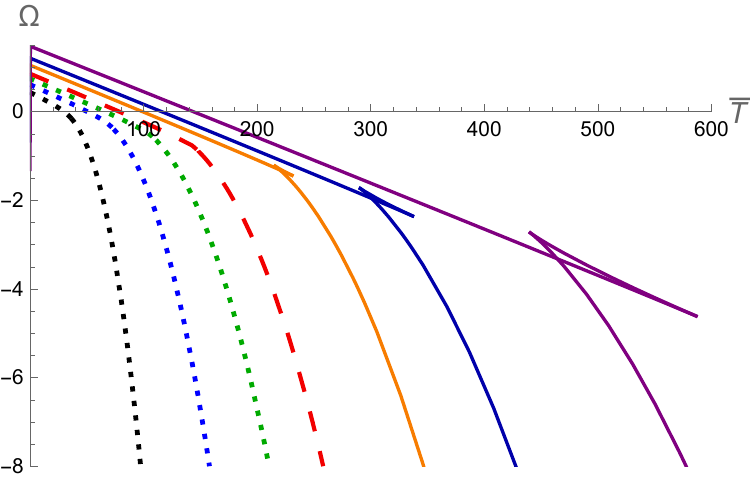}
	\caption{\footnotesize Free energy as a function of temperature for the mixed dyonic nut AdS solutions for $n = 0.2$ and $\phi_e = 1.1$. The dashed line for $\Bar{P}=\Bar{P}_c$. Curves from left to right represent $\Bar{P}=.25 \Bar{P}_c, \Bar{P}=.5 \Bar{P}_c, 0.75 \Bar{P}_c, \Bar{P}_c, 1.5 \Bar{P}_c, 2 \Bar{P}_c$ and $3 \Bar{P}_c$.}
 	\label {mixEunq}
\end{figure}

The free energy diagram shows the phase transition as presented in Fig. \ref{mixEunq}. The Figure indicates that the free energy has a monotonic behavior for all pressures $\Bar{P} \le \Bar{P}_c$, meaning a continuous transition between the small and the large solution radii. Increasing the pressure beyond the critical pressure leads to the appearance of a swallowtail behavior characterizing the first-order phase transition, which is a result of the abrupt change in the entropy of the system.

\begin{figure} [h!]
	\centering
		\includegraphics[width=0.5 \textwidth] {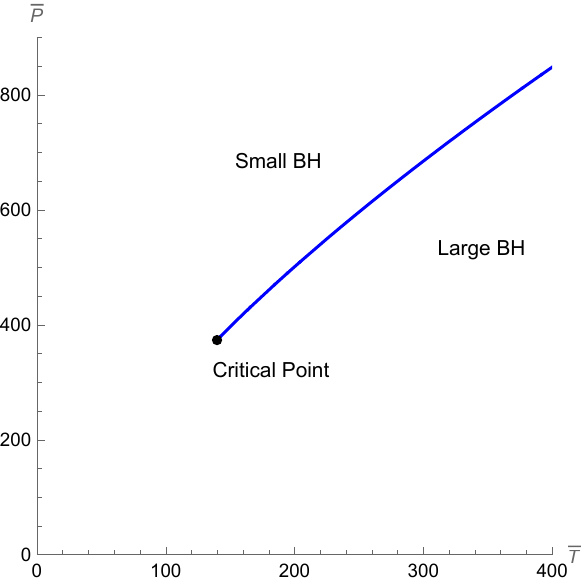}
	\caption{\footnotesize Phase diagram of the mixed dyonic Taub-Nut-AdS solutions for $n = 0.2$ and $\phi_e=1.1$ indicating a continuous transition for $\Bar{P}\le \Bar{P}_c$, while a first-order transition from small to large solutions for $\Bar{P}>\Bar{P}_c$. The critical point $(\Bar{T}_c, \Bar{P}_c)$ is declared by a dot.} 
 	\label {mixphsunq}
\end{figure}

We then expect a phase diagram with a first-order phase transition that begins immediately after $\Bar{P}_c$ and continues endlessly, while for all values of the pressure with $\Bar{P} \le \Bar{P}_c$ the system goes to the one-phase state, or a continuous transition. This is declared in Fig. \ref{mixphsunq}.

%%%%%%%%%%%%%%%%%%%%%%%%%%%%%%%%%%%%%%%%%%%%%%%%%%%%%%%%%%%%%

\subsubsection{Case IV: $\phi_e = 1$}
This case is very similar to the previous one, where again there exists a single critical point. For $n=0.1$ we get the single critical point
\begin{equation}
    \Bar{P}_c = 1200 \,\, , \,\,\,\,\, \Bar{T}_c = 226.27 \,\, , \,\,\,\,\, r_{0(c)} = 0.07 .
\end{equation}
The corresponding AdS scale is 
\be L_c = 0.05 . \ee 
\begin{figure}  [h!]
	\centering
		\includegraphics[width=0.45 \textwidth] {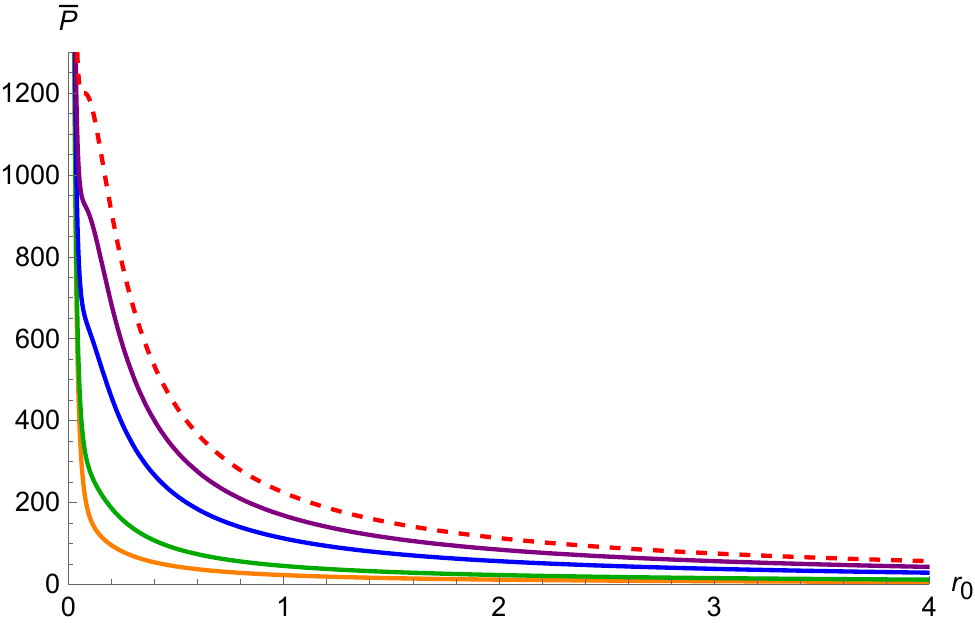}
             \includegraphics[width=0.45 \textwidth] {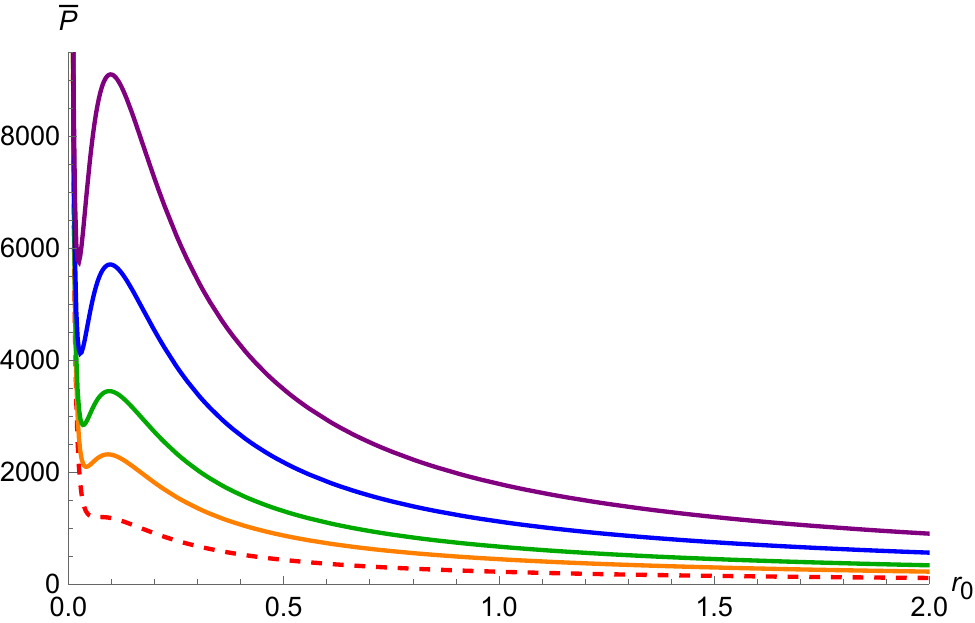}
             \includegraphics[width=0.55 \textwidth] {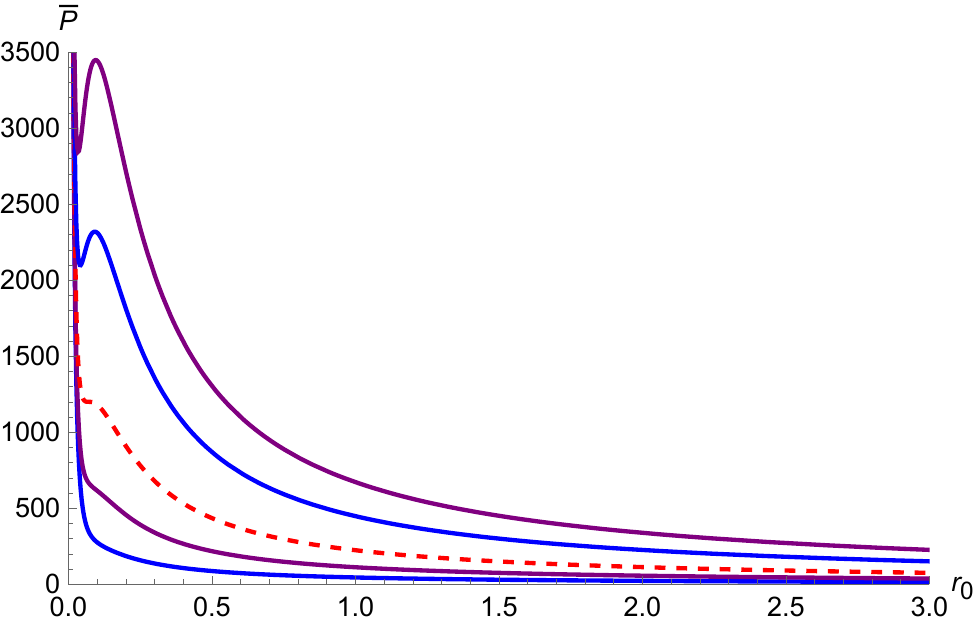}
    	\caption{\footnotesize Mixed dyonic nut solutions showing single critical point behavior for $\phi_e = 1$. Isotherms decrease the temperature from top to bottom for each panel. In calculations $n = 0.1$. The dashed line corresponds to a critical isotherm. \textit{Top left}: Continuous transition for $\Bar{T} \le \Bar{T}_c$. Isotherms from below corresponding to $\Bar{T}=.1 \Bar{T}_c, 0.2 \Bar{T}_c, 0.5 \Bar{T}_c, 0.75 \Bar{T}_c$, and $\Bar{T}_c$. \textit{Top right}: First-order phase transition behavior for $\Bar{T} > \Bar{T}_c$. Isotherms from below: $\Bar{T}= \Bar{T}_c, 2 \Bar{T}_c, 3 \Bar{T}_c, 5 \Bar{T}_c$, and $8 \Bar{T}_c$. \textit{Bottom}: Single critical point behavior for $\phi_e = 1$. Isotherms resembling Van der Waals fluid behavior for $\Bar{T} > \Bar{T}_c$ only. A continuous transition occurs for $\Bar{T} \le \Bar{T}_c$. Isotherms from below: $\Bar{T}=.2 \Bar{T}_c, 0.5 \Bar{T}_c, \Bar{T}_c, 2 \Bar{T}_c$, and $3 \Bar{T}_c$.}
 	\label {mix-unq-1}
\end{figure}

The resulting critical point has the mirror image behavior as in the previous case. Isotherms resemble the Van der Waals fluid when $\Bar{T} > \Bar{T}_c$. Only in these temperature regions we can find a first-order phase transition, see the top right panel of Fig. \ref{mix-unq-1}. While for the temperatures $\Bar{T} \le \Bar{T}_c$, we get a continuous transition in this region, top left panel of Fig. \ref{mix-unq-1}. The temperature $\Bar{T}_0$ is also negative in this region and is given by
\begin{equation}
    \Bar{T}_0 = - \frac{n^2}{r_0^3}  \label{T0_1}
\end{equation}

\begin{figure} [t!]
	\centering
		\includegraphics[width=0.5 \textwidth]{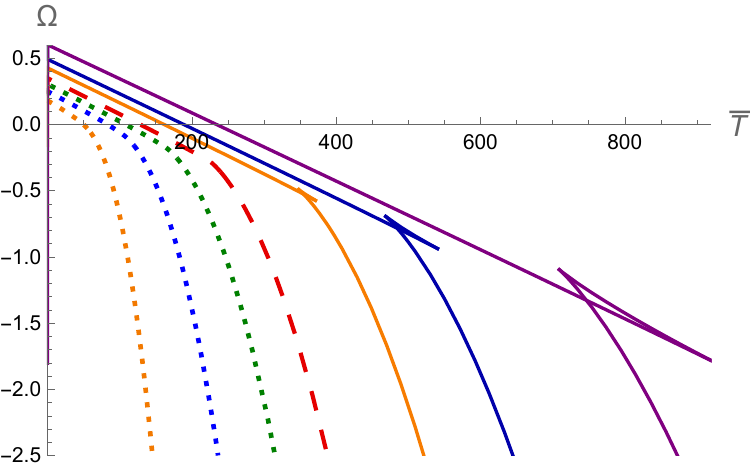} % {mix_phi_1.pdf}
	\caption{\footnotesize Free energy diagram for the mixed dyonic nut AdS solutions for $\phi_e = 1$ at $n = 0.1$ showing a first-order phase transition only for $\Bar{P}>\Bar{P}_c$. The dashed line for $\Bar{P}=\Bar{P}_c$. Curves from left to right represent $\Bar{P}=.25 \Bar{P}_c, \Bar{P}=.5 \Bar{P}_c, 0.75 \Bar{P}_c, \Bar{P}_c, 1.5 \Bar{P}_c, 2 \Bar{P}_c$ and $3 \Bar{P}_c$.}
 	\label {mixE-phi-1}
\end{figure}
\begin{figure} [H]
	\centering
		\includegraphics[width=0.5 \textwidth] {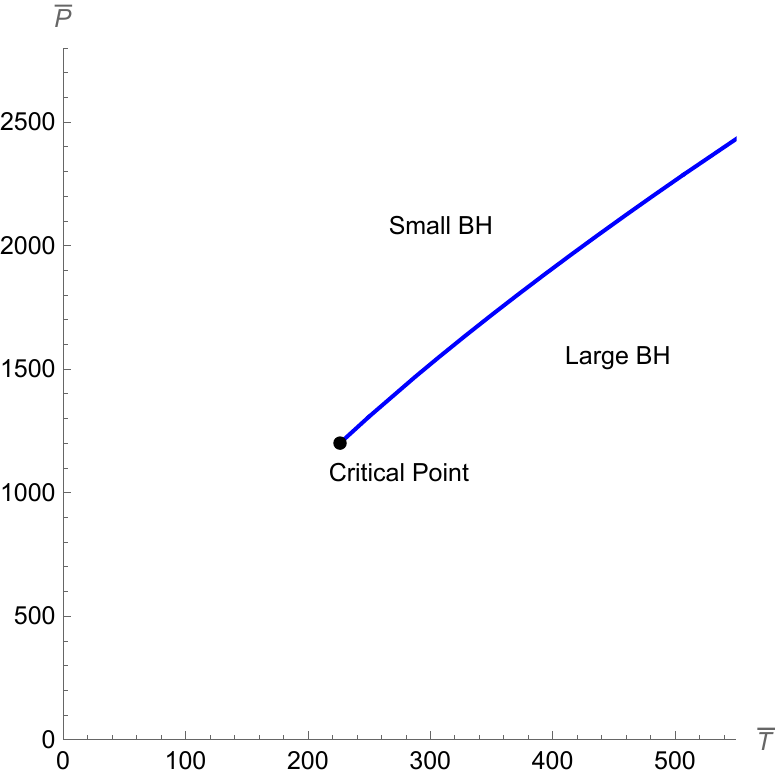}
	\caption{\footnotesize Phase diagram of the mixed Taub-Nut AdS solutions for $\phi_e = 1$ and $n = 0.1$ showing endless first-order small-large radii transition for $\Bar{P}>\Bar{P}_c$. The critical point $(\Bar{T}_c, \Bar{P}_c)$ is declared by a dot.} 
 	\label {mixphs-phi-1}
\end{figure}

The above behavior is confirmed by the free energy diagram, Fig. \ref{mixE-phi-1}, and the phase diagram, Fig. \ref{mixphs-phi-1}.

%%%%%%%%%%%%%%%%%%%%%%%%%%%%%%%%%%%%%%%%%%%%%%%%%%%%%%%%%%%%%%
%\subsection{Nutty Dyons in AdS Space}

\section{Conclusion}

%%%%%%%%%%%%%%%%%%%%%%%%%%%%%%%%%%%%%%%%%%%%%%%%%%%%%%%%%%%%%
%\begin{equation} \Omega = \frac{L^2 \left(\frac{\left(v \left(n^2+r_0)^2\right)+n q_m\right){}^2}{r_0^2}+n^2 \left(2 v^2-1\right)-2 r_0^2 v^2+ r_0^2\right)+ 4 L^2 n q_m v+3 L^2 q_m^2-3 n^4-r_0^4}{4 L^2 r_0} \label{freeE} \end{equation}
In this article, we extend our earlier approach \cite{adel_1,adel_2} to study Taub-NUT-AdS and Dyonic-NUT-AdS solutions by introducing a nut charge $N=n\,(1+4n^2/L^2)$ and its chemical potential $\phi_N$ to the thermodynamics of these solutions. The charge $N$ is conserved since it is the dual quantity of the mass obtained from Komar's integral. One can work as well with the conjugate pair $(n,\phi_n)$ instead of $(N,\phi_N)$, since $\phi_n$ is related to $\phi_N$ by a simple factor. We have shown that in extended thermodynamics, i.e., as we allow AdS radius to change, the enthalpy for Taub-NUT-AdS is $H=M-n\phi_n$. Here the enthalpy is not identified with the gravitational mass anymore, instead, it is related to the mass by a Legendre transform which vanishes as we send $n$ to zero. We have extended this previous idea to the charged dyonic Taub-NUT-AdS solutions to construct a consistent thermodynamics for the dyonic solution. We were able to show that the first law, Gibbs-Duhem, and Smarr's relations are all satisfied. Also, the entropy is the area of the horizon and the temperature goes to that of a dyonic-AdS black hole as $n\rightarrow 0$. An important part of this work is our study of possible phase structures which was analyzed in details using the above approach. We have classified these phase structures into canonical ensembles in which the electric potential is set $\phi_e=0$, and  mixed ones in which the magnetic charge is set $q_m=0$. Our analysis shows some new interesting features which were not reported elsewhere. We found that the phase structure of these solutions is characterized by two distinguished critical points between them there exists a continuous phase transition, especially, in the canonical case and the mixed case with $1/2 \le \phi_e < 1$. We also studied the possibility of merging these two points into one for the canonical and mixed ensembles. Another intriguing cases are those with $\phi_e \ge 1$ in the mixed ensemble which have one critical point but the continuous phase transition region in the $P-T$ diagram is close to the origin in contrast to the usual case of the charged black holes in AdS. Also, the continuous phase transition happens, in this case, if we go to low enough pressure and temperature. It is interesting to check if the two critical point phase structure is analogous to any known fluids in condensed matter systems. This should be interesting, especially upon studying what happens around the merged point! A natural extension of this work is Kerr-NUT-AdS and Kerr-NUT-Newman-AdS which we hope to report on them in the near future.


\begin{thebibliography}{widest entry}
\bibitem{taub} A. H. Taub, Annals of Mathematics 53 (1951) 472.
\bibitem{NUT} E. Newman, L. Tamburino, and T. Unti, Journal of Mathematical Physics 4 (1963), no. 7 915.
\bibitem{misner} C. W. Misner, Journal of Mathematical Physics 4 (1963), no. 7 924.
\bibitem{page+hunter+hawking}S.W. Hawking, C.J. Hunter and D. N. Page, Phys. Rev.{\bf D59}044033 (1999), hep-th/9809035.
\bibitem{hunter+hawking} S.W. Hawking and C.J. Hunter, Phys.Rev. {\bf D59}044025 (1999), hep-th/9808085.
\bibitem{hunter}C.J. Hunter, Phys.Rev.{\bf D59}024009 (1998), gr-qc/9807010.
\bibitem{clifford98}A. Chamblin, R. Emparan, C. V. Johnson, R. C. Myers, Phys.Rev. {\bf D59} 064010 (1999),hep-th/9808177.
\bibitem{bais}S. Bais and P. Batenberg, Nucl. Phys. {\bf B253} 162 (1985)
\bibitem{page+pope}D. N. Page and C. N. Pope, Class. Quant. Grav. {\bf 4} 213 (1987).
\bibitem{adel+andrew}A. Awad and A. Chamblin, Class. Quant. Grav. {\bf 19} 2051 (2002).
\bibitem{adel}A. Awad, Class. Quantum Grav. 23 (2006) 2849.
\bibitem{mann+stelea} R. Mann and Cristina Stelea, Phys. Lett. B 632 (2006) 537; hep-th/0508186.

\bibitem{Bonner} W. B. Bonner, Proc. Camb. Soc. Phil., 66, 145 (1969).
\bibitem{Dowker} J. S. Dowker, General Relativity and Geavitation, Vol. 5, No. 5, 603 (1974).
\bibitem{Ortin-book} T. Ortin, Gravity and Strings, Cambridge University Press (2004).
\bibitem{Kruskal} J. Miller, M. D. Kruskal, and B. B. Godfrey, Physical Review D 4 (1971), no. 10 2945.
\bibitem{Clement1} G. Clement, D. Galtsov, and M. Guenouche, Phys. Lett. B750 (2015) 591, [arXiv:1508.07622].
\bibitem{Clement2} G. Clement, D. Galtsov, and M. Guenouche, Phys. Rev. D93 024048 (2016).
\bibitem{Manko} V. Manko and E. Ruiz, Classical and Quantum Gravity 22 (2005), no. 17 3555.
\bibitem{mann_1} Robie A. Hennigar, David Kubiznak, Robert B. Mann, Phys. Rev. {\bf D100} 064055 (2019)
\bibitem{Geom_inter} A. B. Bordo, F. Gray, R. A. Hennigar, D. Kubiznak, Class. Quant. Grav. 36 (2019) 19, 194001 • e-Print: 1905.03785 [hep-th]
\bibitem{Durka} R. Durka, Int. J. Mod. Phys. D 31 (2022) 04, 2250021; e-Print: 1908.04238 [gr-qc]
\bibitem{FL_rot._taub_nut} A. B. Bordo, F. Gray, R. A. Hennigar, D. Kubiznak. arXiv:1905.06350.
\bibitem{Nutty_dyons_1} A. B. Bordo, F. Gray, and D. Kubiznak, Journal of High Energy Physics 2019 (7) (2019) 119.
\bibitem{nutty_rot} A. B. Bordo, F. Gray, D. Kubiznák, JHEP 05 084 (2020); 2003.02268 [hep-th].
\bibitem{rot_nuts} A. B. Bordo, F. Gray, and D. Kubiznak, Phys.Lett.B 798 (2019) 134972; e-Print: 1905.06350 [hep-th].
\bibitem{nutty_dyons_2} N. Abbasvandi, M. Tavakoli, and R. B. Mann, JHEP 08 152 (2021): 2107.00182 [hep-th].
\bibitem{rot_tn} Ernesto Frodden and Diego Hidalgo, The First Law for the Lorentzian Rotating Taub-NUT; e-Print: 2109.07715 [hep-th].
\bibitem{adel_1} Adel Awad and Somaya Eissa, Phys.Rev. D 101 124011 (2020); e-print: 2007.10489 [gr-qc].
\bibitem{adel_2} Adel Awad and Somaya Eissa, Phys. Rev. D 105, 124034 (2022);e-print: 2206.09124 [hep-th].
\bibitem{clifford_cata} Andrew Chamblin, Roberto Emparan, Clifford V. Johnson, Robert C. Myers, Phys.Rev.D 60 (1999) 064018; e-Print: hep-th/9902170 [hep-th].
\bibitem{smarr} Z. Chen and J. Jiang, Phys. Rev. D {\bf 100}, (2019) 104016; arXiv:1910.10107 [hep-th].
\bibitem{ww} S.~Q.~Wu and D.~Wu, Phys.\ Rev.\ D {\bf 100}, no. 10, (2019) 101501; arXiv:1909.07776 [hep-th].
\bibitem{kastor} D. Kastor, S. Ray, and J. Traschen, Class.Quant.Grav. 26 (2009) 195011, arXiv:0904.2765 [hep-th].
\bibitem{P-V} D. Kubiznak and R. B. Mann, JHEP 07 (2012) 033; e-Print: 1205.0559 [hep-th].
\bibitem{bala} V. Balasubramanian, P. Kraus, Commun.Math.Phys. 208 (1999) 413-428;
e-Print: hep-th/9902121 [hep-th].
\bibitem{surf-ch}  E. Frodden and D. Hidalgo, “Surface Charges Toolkit for Gravity,” Int. J. Mod. Phys. D 29, no.06, 2050040 (2020)
\bibitem{hawking+ross} S.W. Hawking, S.F. Ross, Phys.Rev.D52:5865-5876,1995.
\bibitem{mann06} D. Astefanesei, R. B. Mann, E. Radu, JHEP 0501:049,2006
\bibitem{surf99} R. Emparan, C. V. Johnson, R. C. Myers,Phys. Rev. D 60, 104001 (1999).
\bibitem{gibbons-perry} G.W. Gibbons and M.J. Perry, Phys. Rev. D22, 313 (1980).
\bibitem{pando} R.B. Mann,  L.A Pando Zayas and J. Park High Energ. Phys. 2021, 39 (2021).
\bibitem{clifford14-1} Clifford Johnson, Class. Quant. Grav. 31 (2014) 23, 235003.
\bibitem{clifford14-2} Clifford Johnson, Class. Quant. Grav. 31 (2014) 225005.
\bibitem{BGHK} A. B. Bordo, F. Gray, R. A. Hennigar and D. Kubiznak; arXiv:1905.03785.

\end{thebibliography}
\end{document}